\documentclass[12pt]{article}
\usepackage[xdvi]{graphicx}
\usepackage{amssymb}

\def\nn{\nonumber}
\newcommand{\del}{\partial}
\newcommand{\ddx}[1]{\frac{\partial}{\partial #1}}

\newcommand{\ssstrut}{{\mbox{\scriptsize$\mathstrut$}}}
\newcommand{\A}{\mathcal{A}}
\newcommand{\e}{\mathsf{e}}
\newcommand{\unity}{{\lefteqn{\mathsf{1}}\hspace*{0.11em}\mathsf{1}}}
\newcommand{\C}{C}
\newcommand{\otimesC}{\otimes}
\newcommand{\vol}{\mathit{vol}}

\newcommand{\psibar}{\overline{\psi}}
\newcommand{\Psibar}{\overline{\Psi}}
\newcommand{\zform}{{\mbox{\scriptsize $0$-form}}}
\newcommand{\Dform}{{\mbox{\scriptsize $D$-form}}}
\newcommand{\tr}{\mathop{\mathrm{tr}}\nolimits}
\newcommand{\Tright}[1]{\overrightarrow{T}_{\!\! #1}}
\newcommand{\Tleft}[1]{\overleftarrow{\;T}_{\!\! #1}}

\author{}
\title{Dirac-K\"ahler Fermion from Clifford Product with
Noncommutative Differential Forms on a Lattice}

\begin{document}
\renewcommand{\theequation}{\arabic {section}.\arabic{equation}}
\setlength{\baselineskip}{7mm}
\begin{titlepage}
\begin{flushright}
EPHOU-03-001 \\
May, 2003
\end{flushright}

\vspace{15mm}
\begin{center} 
{\Large Dirac-K\"ahler Fermion from Clifford Product with
Noncommutative Differential Form on a Lattice}

\vspace{1cm}

{\sc Issaku Kanamori}\footnote{kanamori@particle.sci.hokudai.ac.jp} and 
{\sc Noboru Kawamoto}\footnote{kawamoto@particle.sci.hokudai.ac.jp} 
 \\
{\it{ Department of Physics, Hokkaido University }}\\
{\it{ Sapporo, 060-0810, Japan}}\\
\end{center}
\vspace{2cm}

\begin{abstract}

We formulate Dirac-K\"ahler fermion action by introducing a new 
Clifford product with noncommutative differential form on a lattice.
Hermiticity of the Dirac-K\"ahler action requires to choose the 
lattice structure having both orientabilities on a link. 
The Kogut-Susskind fermion and the staggered fermion actions are 
derived directly from the Dirac-K\"ahler fermion formulated by 
the Clifford product.  
The lattice QCD action with Dirac-K\"ahler matter fermion is also 
derived via an inner product defined by the Clifford product. 
\end{abstract}

\vspace{5 cm}

\begin{tabular}{lcl}
\small
 PACS code &:& 04.60.Nc, 11.15.Ha \\
 Keywords &:&  Dirac-K\"ahler fermion, noncommutative geometry, 
 Clifford product, \\ 
          &&lattice fermion, lattice gauge theory
\end{tabular}

\end{titlepage}


\section{Introduction}
\setcounter{equation}{0}

One of the fundamental questions of the modern field theory is obviously 
to find a possible origin of supersymmetry if any. 
It was pointed out by Witten that the 
quantization of the topological Yang-Mills action with twisting mechanism 
in 4 dimensions generates N=2 super Yang-Mills action\cite{Witten1}. 
It was later recognized that this quantization is realized essentially 
by the instanton gauge fixing of Yang-Mills field\cite{B-S,L-P,B-M-S}. 
It is thus natural to expect that the twisting mechanism may play 
an important role to understand the possible origin of supersymmetry 
at least in the special case like topological field theory. 

One of the authors (N.K.) and Tsukioka showed that the 2-dimensional 
example of a generalized topological Yang-Mills action 
of generalized gauge theory \cite{Kawamoto:1992xq} with instanton 
gauge fixing generates 2-dimensional super Yang-Mills action 
in the exactly similar way as the 4-dimensional case\cite{Kawamoto:1999zn}. 
They claimed that the twisting mechanism is just Dirac-K\"ahler fermion 
mechanism\cite{I-L,Kahler:1962,Graf:1978kr}\cite{Becher:1982ud,Banks:1982iq,
Rabin:1982qj,Benn:1983sr,Mitra:ci,
Aratyn:tq,Bullinaria:1986ed,Solodukhin:1991dj,Kruglov:2001wd} 
which is formulated by 
differential forms. 
Recently a twisted version of superspace formulation 
has been developed and recognized to play a crucial role in this formulation 
of supersymmetry and Dirac-K\"ahler fermion formulation\cite{K-K-U}. 
It was recognized that the \lq\lq flavor'' suffix of Dirac-K\"ahler fermion 
is equivalent to the extended supersymmetry suffix. 
It is thus very natural to expect that Dirac-K\"ahler fermion formulation 
plays an important role in formulating the field theory of extended 
supersymmetry. 

On the other hand it was recognized for a long time that the Dirac-K\"ahler 
fermion, the Kogut-Susskind fermion and the staggered fermion on the lattice 
are closely related and in fact almost 
equivalent
\cite{Kogut:1975ag,Susskind:1977jm}
\cite{Kawamoto:1981hw,Kluberg-Stern:1983dg,Gliozzi:1982ib}. 
The Dirac-K\"ahler fermion can be formulated by Clifford product of 
differential form and then the 
one to one correspondence of $n$-form of differential algebra and 
$n$-simplex of the lattice structure is essential in the identification of 
the Dirac-K\"ahler fermion formulation and the lattice fermion formulations. 
In order to define Dirac-K\"ahler fermion formulation on the lattice 
we need to formulate differential form and then Clifford product on the 
lattice. 

There are several fundamental difficulties in formulating supersymmetry 
on the lattice. The most notorious problem is the lattice chiral fermion 
problem. Here we would like to consider that the breakdown of Leibniz 
rule for the differential operator on the lattice is another fundamental 
difficulty. 
To be more explicit it is the well-known fact that the differential 
operator defined as a difference operator on the lattice does not satisfy 
Leibniz rule:
\begin{equation}
 d\{f(x)g(x)\} \neq d\{f(x)\}g(x) +  f(x)d\{g(x)\},
\end{equation} 
where we define 
\begin{equation}
 d\{f(x)\} = \partial_{+\mu}f(x) dx^\mu = 
            \{f(x+\mu)-f(x)\}dx^\mu.
\end{equation} 
If we, however, introduce the following noncommutativity between 
the function and differential:
\begin{equation}
 dx^\mu f(x) = f(x+\mu)dx^\mu,
\end{equation} 
the Leibniz rule on the lattice can be fulfilled:
\begin{eqnarray}
 d\{f(x)g(x)\} 
               = d\{f(x)\}g(x) +  f(x)d\{g(x)\}.
\end{eqnarray} 
This type of noncommutative differential algebra formulated especially on the 
lattice was developed by several authors\cite{Woronowicz:1989rt,
Dimakis:1992pk,Dimakis:1993pj,Aschieri:1993wg,Dimakis:1994qq,Dimakis:1994bq,
Aschieri:2002vn}. 
In formulating supersymmetry on the lattice Leibniz rule may be one of the 
most important characteristics to be satisfied. 
This formulation with noncommutative differential forms is 
a special case of the noncommutative geometry {\it {\`a} la} 
Connes\cite{Connes:1994-text,Landi:1997sh} but not the same type of 
noncommutativity proposed in 
string theory\cite{Seiberg-Witten}. 

In this paper we formulate Dirac-K\"ahler fermion formulation 
on a lattice with noncommutative differential forms as a first step of 
finding a formulation of supersymmetric field theory on a lattice. 
We derive the Kogut-Susskind fermion and the staggered fermion actions 
directly from the Dirac-K\"ahler fermion action. 
We try to identify the origin of the difference in the Kogut-Susskind 
fermion and the staggered fermion formulations. 
In formulating Dirac-K\"ahler fermion we need to introduce Clifford 
product on the lattice, which actually provides more general framework 
of formulating lattice theory including the lattice gauge theory 
with noncommutative differential forms which has already been developed by 
several authors\cite{Dimakis:1992pk,Dimakis:1994qq,Aschieri:2002vn,
Balachandran:1998qt,Dai:2001td}. 
In defining the Clifford product with noncommutative differential forms 
on the lattice the proof of associativity is crucial. We have found a 
definition of associative Clifford product. 
A simpler version of the definition of Clifford product with noncommutative 
differential form on the lattice was presented in \cite{kanamori1}.

There was an interesting but ad hoc derivation of the staggered fermion 
from the formulation of noncommutative differential form by J. Dai and 
X. -C. Song\cite{Dai:2000vf,Dai:2001wn}. 
In their derivation, however, the relation between the staggered fermion 
and the Dirac-K\"ahler fermion formulations is not clear.  
We have derived the Kogut-Susskind 
fermion and the staggered fermion directly from Dirac-K\"ahler fermion 
formulation by a newly defined Clifford product. 
The general definition of Clifford algebra on the lattice was 
investigated by Vaz, where 1-form Clifford product was explicitly 
given\cite{Vaz1,Vaz2}. 
There are some comments 
on the Dirac-K\"ahler formulation and noncommutative differential 
form in the literature\cite{Striker:1995vg}.  

Another important motivation of formulating Dirac-K\"ahler fermion 
formulation with noncommutative differential form is an introduction 
of matter fermion into lattice gravity. 
The 2-dimensional quantum gravity was successfully formulated by dynamical
triangulation of 2-dimensional space time\cite{BKKM,ADJ2}. 
The introduction of matter fermion on the triangulated 
lattice was successfully formulated. 
It was shown that the essence of the 
2-dimensional quantum gravity on the lattice is the fractal structure 
of the 2-dimensional space time\cite{KKSW} 
and the fractal dimensions were 
evaluated numerically and analytically and perfectly consistent
\cite{KKSW,KWY1,KWY2}\cite{DHK,KN,KSW,KKMW}\cite{K-Y}.  
The topological versions of Chern-Simons gravity in 3-dimensions and 
BF gravity in 4-dimensions were successfully formulated on a lattice
\cite{K-N-S,K-S-U}. 
It is important to find 
a formulation of introducing matter fermion on these pure 
gravity on the simplicial lattice 
manifold. If a matter fermion like Dirac-K\"ahler fermion is formulated by 
the differential form it will provide a coordinate independent 
formulation and thus may lead to a lattice gravity formulation 
with matter. 

It is interesting to point out that the Weinberg-Salam model was formulated 
by using the generalized gauge theory\cite{Kawamoto:1992xq} with 
a graded Lie algebra of super group $SU(2\mid 1)$\cite{K-T-U}.
In this formulation the Clifford product was crucial in defining 
Yang-Mills action of generalized gauge theory and the introduction 
of Dirac-K\"ahler matter fermion. 
There is a speculative hope that the standard model and gravity could be 
formulated on the simplicial lattice in terms of differential forms. 

This paper is organized as follows: In section 2 we summarize the 
formulation of noncommutative differential forms on a lattice in particular 
the basic formulation of universal differential algebra on a lattice.  
We point out the fundamental connection of Hermiticity and lattice structure. 
We summarize the known results of Dirac-K\"ahler fermion formulation and the 
staggered fermion formulation in section 3. 
We formulate Dirac-K\"ahler fermion from Clifford product on a oriented 
lattice in section 4. We propose new Clifford products on a 
symmetric lattice and discuss the connection between Hermiticity and the 
positivity of self inner product of a field in section 5.   
In section 6 we formulate Dirac-K\"ahler fermion and the staggered fermion 
by using the Clifford product defined on the symmetric lattice. 
In section 7 we formulate the lattice gauge theory with Dirac-K\"ahler 
matter fermion in terms of the Clifford product of noncommutative 
differential forms. In section 8 we give conclusion and discussions.


\section{Noncommutative Differential Form on a Lattice}

\setcounter{equation}{0}

We first summarize the formulation of noncommutative differential forms on a
lattice.
To define differential forms on a discrete space, we need some formal
description.
We treat the lattice as discrete Abelian group. 
See for example \cite{Dimakis:1994bq} and 
\cite{Aschieri:1993wg,Aschieri:2002vn} for more general treatment.

\subsection{Universal Differential Algebra on a Lattice}

A function on a lattice is a map from lattice point $x$ to a complex
number.
These functions constitute an algebra which is
associative and Abelian, and possesses unity.
We call the algebra $\A$.
We can expand a function $f \in \A$ as
\begin{equation}
 f = \sum_x f_x \e^x,
\label{function-of-differential-algebra} 
\end{equation} 
where $f_x \in \C$ is a coefficient and $\e^x \in \A$ spans a basis
of $\A$. 
The sum $\sum_x$ runs over all the lattice points $x$. 
$\e^x$ has the following properties:
\begin{eqnarray}
 \e^x (y) = \delta^{xy} \label{eq:ex}, \\
 \e^x \e^y = \delta^{xy}\e^x \label{eq:exey}, \\
 \sum_x \e^x = \unity .
  \label{eq:e-completeness}
\end{eqnarray}
The relation (\ref{eq:ex}) indicates that $\e^x$ plays a role of delta 
function when acting on a lattice point $y$. 
It should thus be noted that 
$f(x) = \sum_y f_y \e^y(x) = f_x$.
The relation (\ref{eq:exey}) means that the product of functions works 
pointwise, and ensures its Abelian nature. 
The relation (\ref{eq:e-completeness}) assures the completeness of the 
bases.

$1$-form is $\A$-bimodule generated by the following differential 
operator $d$:
\begin{eqnarray}
  & \displaystyle d(fg) = (df)g + f\, dg  \qquad\qquad &
       \big( \mbox{Leibniz rule}\big), \label{eq:d_def1}\\
  & \displaystyle d( \alpha f + \beta g) = \alpha\,df + \beta\, dg \qquad\qquad
        & \big( \mbox{linearity}\big), \label{eq:d_def2}
\end{eqnarray}
where $f,g \in \A$ and $\alpha, \beta \in \C$. $\A$-bimodule is the 
direct sum of left $\A$-module and right $\A$-module, where 
the left $\A$-module has coefficient functions on the left of 
the element of the algebra $\A$ and the right ${\cal A}$-module 
has the similar form with left and right interchanged.
The Leibniz rule makes it possible that we can write any $1$-form as
\begin{equation}
 \omega = \sum_i f_i\,dg_i,
\end{equation}
which has the form of left $\A$-module.
We can check that the following representation of $d$
\begin{equation}
 df = \unity \otimes f - f \otimes \unity \label{eq:df}
\end{equation}
satisfies the above two properties (\ref{eq:d_def1}) and (\ref{eq:d_def2}).
Here $\unity$ is defined in (\ref{eq:e-completeness}).

Using equation (\ref{eq:df})
we obtain the differential of the basis $\e^x$
\begin{equation}
 d \e^x = \sum_y \left( \e^{y,x} - \e^{x,y}\right),
\end{equation}
where 
\begin{equation}
 \e^{x,y} \equiv \left\{
	      \begin{array}{@{\,}l@{\qquad}l}
		 \e^x\otimes \e^y \ ( = \e^x \,d\e^y) & (x \neq y) \\
		 0                 & (x=y).
	      \end{array}
			 \right.
\end{equation}
$\e^{x,y}$ is defined on the link which connects the sites $x$ and $y$. 
For simplicity we may simply call the basis of link variable $\e^{x,y}$ as 
the link connecting the sites $x$ and $y$. 

We define the following differential corresponding to $1$-form:
\begin{equation}
 \theta^x = \sum_y \e^{y,y+x}, 
\label{eq:def-1-form}
\end{equation}
which is shift invariant in the following sense:
\begin{equation}
 \sum_y \e^{y,y+x} = \sum_y \e^{y+z,y+z+x}.
\end{equation}
Using this quantity, we can rewrite the differential of a function as
\begin{equation}
 df = \sum_{x,y} f_x \left( \e^{y,x} - \e^{x,y}\right)
    = \sum_x (T_x f - f) \theta^x, \label{eq:universal-df}
\end{equation}
where shift operator $T_x$ acts as
\begin{equation}
 T_x f = T_x \sum_y f_y \e^y = \sum_y f_{y+x} \e^y = \sum_y f_y \e^{x-y}.
\end{equation}

It is important to realize the noncommutativity of functions and 
$1$-forms,
\begin{eqnarray}
 \e^x \theta^y = \theta^y \e^{x+y}, \nonumber \\
 \theta^x f = (T_x f)\theta^x,  
 \label{eq:noncommutativity}
\end{eqnarray}
while $\e^x \e^{y,z} \neq \e^{y,z}\e^x$ in general.

We next introduce higher forms.
We first define the differentiation of $\e^{x,y} = \e^x d\e^y (x \ne y)$ by 
\begin{equation}
 d(\e^{x,y})  = \sum_z\left( \e^{z,x,y} - \e^{x,z,y} + \e^{x,y,z}\right),
 \label{eq:d-for-1form}
\end{equation}
where 
\begin{equation}
   \e^{x,y,z}
    \equiv \e^{x,y} \e^{y,z}
    =  \left\{\begin{array}{@{\,}l@{\qquad}l}
	\e^x \otimesC \e^y \otimesC \e^z &
             \big( y \neq x \mbox{ and } y \neq z\big) \\
	0       &  \big( \mbox{otherwise} \big).\\
	      \end{array}\right.
\end{equation}
This definition assures the nilpotency of the differential operator $d$, 
\begin{equation}
 d^2 \e^x = 0,
\end{equation} 
and thus leads
\begin{equation}
 d(\e^{x,y})  = d\e^x d\e^y
\end{equation}
which defines the $d$ action on $1$-form, and ensures the Leibniz rule. 
$\e^{x,y,z}$ spans the basis of $2$-form and can be rewritten in the 
following form for $x\neq y$ and $y\neq z$:
\begin{equation}
 \e^{x,y,z} = \e^x\,d\e^y\,d\e^z = \e^x \theta^{y-x} \theta^{z-x-y}.
\end{equation} 
The pointwise nature of the product leads the following relations:
\begin{eqnarray}
   \e^x \e^{y,z} = \delta^{xy} \e^{y,z},
    \qquad
   \e^{x,y} \e^z = \delta^{yz} \e^{x,y} \nonumber\\
   \e^x \e^{y,z,w} = \delta^{xy} \e^{y,z,w},
    \qquad
   \e^{x,y,z} \e^w = \delta^{zw} \e^{x,y,z}
 \label{eq:ex-eyzw}
\end{eqnarray}
and
\begin{equation}
    \e^{x,y} \e^{z,w}
        = \delta^{y,z} \e^{x,y} \e^{z,w} = \delta^{y,z}\e^{x,y,w}.
  \label{eq:exy-ezw}
\end{equation}

Generalization to higher form is straightforward. 
A basis of $p$-form is
\begin{eqnarray}
    \lefteqn{\e^{x_0,x_1,\ldots,x_p}}\hspace{1em}\nonumber\\
     &\equiv& \e^{x_0,x_1} \e^{x_1,x_2} \ldots \e^{x_{p-1},x_p} \\
     &=& \left\{\begin{array}{@{\,}l@{\qquad}l}
	   \e^{x_0} d\e^{x_1} d\e^{x_2} \ldots d\e^{x_p}
              & \big( x_0 \neq x_1 \mbox{ and } x_1 \neq x_2 \mbox{ and }
	        \ldots x_{p-1} \neq x_p \big)\\
           0  & \big( \mbox{otherwise} \big),
		\end{array}\right.\nn
\end{eqnarray}
and $d$ acts as
\begin{equation}
    d \e^{x_0,\ldots ,x_p}
     = \sum_y\sum_{q=0}^{p+1} (-)^q \e^{x_0,\ldots, x_{q-1} ,y,x_q,\ldots,x_p}.
     \label{eq:d-for-e}
\end{equation}
Similar relations as (\ref{eq:ex-eyzw}) and (\ref{eq:exy-ezw}) hold for
$p$-forms as well.
A general $p$-form is
\begin{equation}
  \omega_p = \sum_{\{x_i\}} f_{x_0 x_1 \ldots x_p} \e^{x_0,x_1,\ldots,x_p },
  \label{eq:def_of_p-form }
\end{equation}
or using $\theta^{x_i}$
\begin{equation}
    \omega_p = \sum_{\{x_i\}} f'_{x_0 \ldots x_p} \e^{x_0}
                \theta^{x_1} \ldots \theta^{x_p}.
\end{equation}
It is straightforward to check the Leibniz rule and nilpotency of the 
differential operator,
\begin{eqnarray}
   &\displaystyle d^2=0 \label{eq:nilpotency}\\
   &\displaystyle
      d(\omega_p \omega_q) = (d\omega_p)\omega_q + (-)^p \omega_p\,d\omega_q,
    \label{eq:Leibniz}
\end{eqnarray}
where $\omega_p$ and $\omega_q$ are $p$-form and $q$-form, respectively.
The algebra of these forms is called universal differential algebra.

\subsection{Hermiticity and Lattice Structure}

The differential 1-form defined in (\ref{eq:def-1-form}) of the universal 
differential algebra contains all possible $\e^{x,y}$ connecting a site 
$x$ to the rest of all lattice sites $y$. 
The 1-form $\e^{y,y+x}$ in the summation is therefore highly non-local. 
In order to obtain the standard lattice formulation of known actions 
we need to make a reduction of links by truncating most of the non-local 
links $\e^{x,y}$ except for the nearest neighboring links. 
We introduce two types of reduction, one is oriented lattice reduction 
which is asymmetric and the other is symmetric lattice reduction.
These types of reduction were proposed by 
Dimakis and M\"uller-Hoissen\cite{Dimakis:1994qq}.

We consider the following two types of square lattice:
\begin{eqnarray}
&1)&\ \ \hbox{oriented lattice (see Fig.1)}: 
       \e^{x,y} \left\{
	   \begin{array} {@{\;}l@{\qquad}l}
	    \neq 0 & \big( y = x+\mu \big) \\
	    = 0 &  \big( \mbox{otherwise} \big) 
	   \end{array} \right.  \\
&2)&\ \ \hbox{symmetric lattice (see Fig.1)}:
       \e^{x,y} \left\{
	   \begin{array} {@{\;}l@{\qquad}l}
	    \neq 0 & \big( y = x \pm \mu \big) \\
	    = 0 &  \big( \mbox{otherwise} \big) .
	   \end{array}\right.
\end{eqnarray}
In the oriented lattice reduction the links pointing from $x$ to $x+\mu$ 
which is next neighbor site of $x$ in the $\mu$ direction survive while 
the links with the opposite direction pointing from $x$ to $x-\mu$ are 
truncated. Thus links have orientability in one direction. 
In the symmetric lattice reduction links with both directions 
$x$ to $x \pm \mu$ survive and thus links are symmetric in the 
orientability. 
Hereafter we differentiate the two cases by 1) and 2) for oriented lattice 
reduction and symmetric lattice reduction, respectively.

\begin{figure}
 \includegraphics[width=\linewidth]{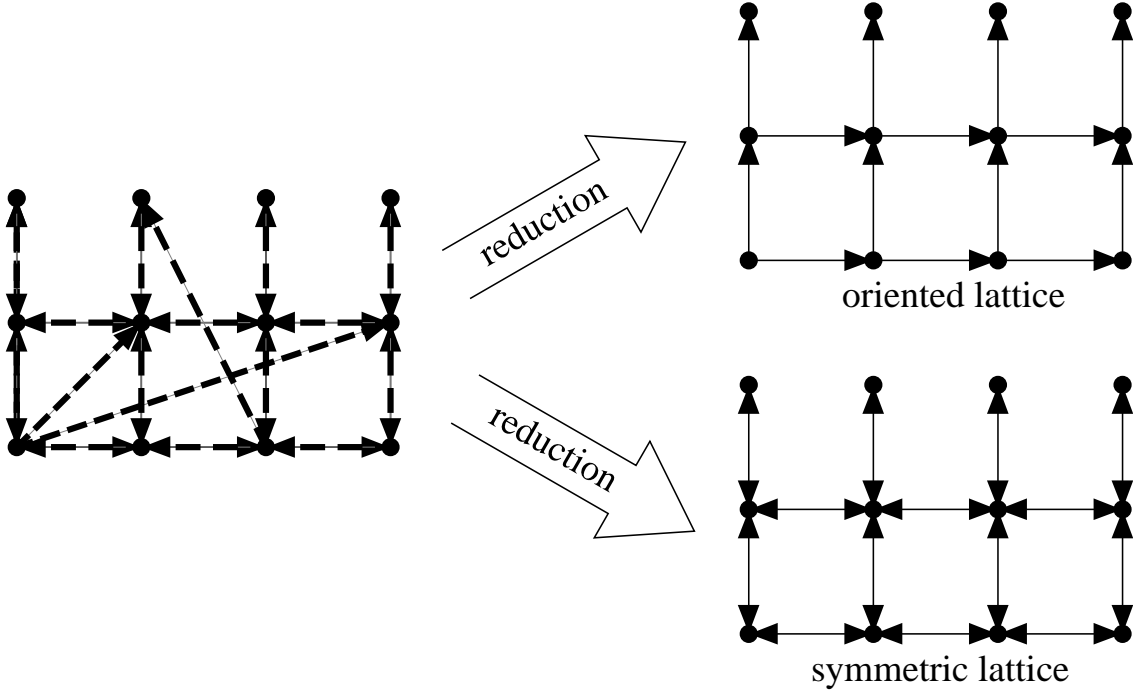}
 \caption{ Possible reductions into 1) oriented lattice, and 
 2) symmetric lattice.}
 \label{fig:reduction-o}
\end{figure} 

For the oriented reduction there are $D$ nonzero $\theta^x$; 
$\theta^\mu = \sum_x \e^{x,x+\mu}$ while for the 
symmetric reduction there are $2D$ nonzero $\theta^x$; 
$\theta^{\pm\mu} = \sum_x \e^{x,x\pm\mu}$. 
The noncommutativity of functions and 1-forms shown in 
(\ref{eq:noncommutativity}) on the lattice can now be given more explicitly 
by
\begin{eqnarray}
 \theta^{\pm\mu} f_x &=& f_{x\pm\mu} \theta^{\pm\mu}. 
 \label{noncommutativity}
\end{eqnarray}
Here the noncommutativity of bases $\e^x$ and $\theta^{\pm\mu}$ 
shown in Eq.(\ref{eq:noncommutativity}) can be reinterpreted as 
the noncommutativity of components and forms.  

The differential defined in (\ref{eq:universal-df}) can be given by 
\begin{eqnarray}
 1) \ df &=& \sum_{\mu=1}^D (T_\mu f - f )\theta^\mu
    = \sum_{\mu=1}^D \del_{+\mu}f \theta^\mu
    = \sum_{\mu=1}^D \theta^\mu \del_{-\mu}f  ,\\
 2) \ df
   &=& \sum_{\mu=1}^{D}
       \left[ (T_\mu f - f) \theta^\mu + (T_{-\mu}f -f )\theta^{-\mu}\right]
      \nn \\
   &=& \sum_{\mu=1}^D
        \left[
          \frac{1}{2}(\del_{+\mu} + \del_{-\mu})f 
          (\theta^{\mu} - \theta^{-\mu})
          + \frac{1}{2}(\del_{+\mu} - \del_{-\mu})f 
          (\theta^{\mu} + \theta^{-\mu})\right]  \nn \\
   &=& \sum_{\mu=1}^D
        \left[
          (\theta^{\mu} - \theta^{-\mu})
          \frac{1}{2}(\del_{+\mu} + \del_{-\mu})f 
          - (\theta^{\mu} + \theta^{-\mu})
          \frac{1}{2}(\del_{+\mu} - \del_{-\mu})f                     
        \right],
       \label{eq:differential0}
\end{eqnarray}
where $\del_{\pm\mu}$ is forward and backward difference operators defined by
\begin{eqnarray}
 \del_{\pm\mu} f_x = \pm(T_{\pm\mu}-1)f_x = \pm(f_{x\pm\mu}-f_x). 
\end{eqnarray}
The differential for the oriented reduction needs only the forward difference 
operator while the differential for the symmetric reduction needs both 
of the forward and backward difference operators.
Operating $d$ on $x^\mu$, we can find a natural identification of 
the differential 1-form on the lattice as 
\begin{equation}
1)\ \ d x^\mu \equiv \theta^\mu, \ \ \ 
2)\ \ dx^\mu \equiv \theta^\mu - \theta^{-\mu}. 
\label{dx-of-2)}
\end{equation}
We, however, need to introduce different type of 1-forms associated with 
the second derivative in eq.(\ref{eq:differential0}) on the lattice 
in the symmetric reduction,   
\begin{equation}
2)\ \ \tau^\mu \equiv \theta^\mu + \theta^{-\mu}.
\label{tau-of-2)}
\end{equation}
For general $p$-forms $d$-operation then leads 
\begin{eqnarray}
 1) \ d\omega_p &=&  \sum_{\mu=1}^D dx^\mu \del_{-\mu}\omega_p,\\
 2) \ d\omega_p &=& \sum_{\mu=1}^D
        \left[
          dx^\mu\frac{1}{2}(\del_{+\mu} + \del_{-\mu}) \omega_p
          - \tau^\mu \frac{1}{2}(\del_{+\mu} - \del_{-\mu})\omega_p
        \right],
       \label{eq:differential}
\end{eqnarray}
where the differentials need proper location.
These definitions of the exterior derivative on the lattice satisfy 
the Leibniz rule
\begin{equation}
d(fg) = (df)g + fdg. 
\end{equation}

For both 1) and 2) 
we have $\e^{x,x+\mu+\nu} = 0$, which leads
\begin{eqnarray}
   0 = d \e^{x,x+\mu+\nu}
     = - \e^{x,x+\mu,x+\mu+\nu} -\e^{x,x+\nu,x+\mu+\nu}
     \qquad (\ \mu \neq \nu \ ).
\end{eqnarray}
Summing $x$ up, we obtain
\begin{equation}
1),\ \ 2)\ \  \ \theta^\mu \theta^\nu + \theta^\nu \theta^\mu = 0.
    \label{eq:extp1}
\end{equation}
Similarly, from $\e^{x,x+\mu+\mu}=0$, we obtain
\begin{equation}
1),\ \ 2)\ \  \ \theta^\mu \theta^\mu = 0
      \qquad \big( \mbox{no summation for $\mu$} \big).
  \label{eq:extp2}
\end{equation}
Equations (\ref{eq:extp1}) and (\ref{eq:extp2}) are just exterior
product algebras.
For the symmetric reduction we further obtain the similar relations, 
\begin{equation}
2)\ \ \theta^{-\mu} \theta^{-\nu} + \theta^{-\nu} \theta^{-\mu} =0,
\end{equation}
from $\e^{x,x-\mu-\nu} = 0$ and
\begin{equation}
2)\ \  \sum_\mu \left[
    \theta^\mu \theta^{-\mu} + \theta^{-\mu} \theta^\mu \right]=0
\end{equation}
from $\e^{x,x}=0$.
Though the last equation holds after summation of $\mu$, 
we impose stronger conditions
\begin{equation}
 2)\ \   \theta^\mu \theta^{-\mu} + \theta^{-\mu} \theta^\mu =0
   \qquad \mbox{( no summation )}
\end{equation}
so that all $\theta^{\pm \mu}$ satisfy exterior algebra.

We next define $\ast$-conjugation so as to satisfy the following first condition:
\begin{equation}
 (d\omega_p)^\ast = d(\omega_p)^\ast ,
 \label{eq:star-condition-1}
\end{equation}
where $\omega_p$ is the $p$-form defined in (\ref{eq:def_of_p-form }) 
and $d$ acts as in (\ref{eq:d-for-e}).
Then there are two possible definitions of the $\ast$-conjugation which are 
consistent with the above condition (\ref{eq:star-condition-1}),
\begin{eqnarray}
&\hbox{A)}&(\e^{x_0,x_1,\ldots,x_p })^\ast = \e^{x_0,x_1,\ldots,x_p }, \\
&\hbox{B)}&(\e^{x_0,x_1,\ldots,x_p })^\ast = 
(-1)^{\frac{p(p+1)}{2}} \e^{x_p,x_{p-1},\ldots,x_0 }.
\end{eqnarray}
We impose the second condition A) or B) to define $\ast$-conjugation. 
There are, however, the following two possible variations of condition: 
\begin{eqnarray}
\hbox{a)}\ \ (\omega_p \omega_q)^\ast = 
\omega_p^\ast \omega_q^\ast, 
\ \ \ 
\hbox{b)}\ \ (\omega_p \omega_q)^\ast = 
(-1)^{pq}\omega_q^\ast \omega_p^\ast, 
\end{eqnarray}
where $\omega_p$ and $\omega_q$ are $p$-form and $q$-form, respectively. 
Here we point out that the choice of the conditions A) and a) is adequate 
for the definition of $\ast$-conjugation on the oriented lattice 1) 
while the choice of B) and b) is adequate for the symmetric lattice 2). 

In these definitions the $\ast$-conjugation changes left ${\cal A}$-module 
into left ${\cal A}$-module on the oriented lattice while left 
${\cal A}$-module to right ${\cal A}$-module on the symmetric lattice, 
\begin{eqnarray}
&\hbox{1)}&
(\omega_{\mu_1,\cdots,\mu_p}\theta^{\mu_1} \cdots \theta^{\mu_p})^\ast = 
(\omega_{\mu_1,\cdots,\mu_p})^* 
(\theta^{\mu_1})^\ast \cdots (\theta^{\mu_p})^\ast \\
&\hbox{2)}&
(\omega_{\epsilon_1\mu_1,\cdots,\epsilon_p\mu_p}
\theta^{\epsilon_1\mu_1} \cdots \theta^{\epsilon_p\mu_p})^\ast = 
(-1)^{\frac{p(p-1)}{2}}(\theta^{\epsilon_p\mu_p})^\ast \cdots 
(\theta^{\epsilon_1\mu_1})^\ast 
\omega_{\epsilon_1\mu_1,\cdots,\epsilon_p\mu_p}^*,
\end{eqnarray}
where $\epsilon_i=\pm$ and 
$\omega_{\epsilon_1\mu_1,\cdots,\epsilon_p\mu_p}^*$ can be taken either as 
a simple complex conjugation or a Hermite conjugate. 
According to the choice A) and B) for 1) and 2), respectively, we obtain 
\begin{eqnarray}
\hbox{1)}\ \ (\theta^{\mu})^\ast = \theta^{\mu}, 
\ \ \ 
\hbox{2)}\ \ (\theta^{\pm\mu})^\ast = -\theta^{\mp\mu}.
\end{eqnarray}
For the symmetric lattice we identify the following interesting features:
\begin{eqnarray}
\hbox{2)}\ \ (dx_{\mu})^\ast = dx_{\mu}, \ \ \ 
(\tau_\mu)^\ast = -\tau_\mu,
\label{real-imaginary-star-relation}
\end{eqnarray}
where $dx_{\mu}$ behaves as real while $\tau_\mu$ behaves as pure imaginary 
with respect to the $\ast$-conjugation. 
It should be noted that the order of the function and the differentials 
are important due to their noncommutativity. 
It then turns out that the $\ast$-conjugation on the oriented lattice 
plays a similar role as complex conjugation while the 
$\ast$-conjugation on the symmetric lattice plays a role of Hermite 
conjugate according to the symmetric nature of the lattice structure.

It is highly nontrivial how to define the integration with noncommutativity 
on the lattice. It is natural to define the integral as the inverse of 
differentiation: 
\begin{equation}
 \int df = f.
\end{equation}
Then in 1-dimension we define the integration on the lattice 
with noncommutativity so as to satisfy 
\begin{equation}
 \int_x^{x+a} df = f \Bigr|_x^{x+a},
 \label{integration1}
\end{equation}
in particular 
\begin{equation}
 \int_x^{x+a} dx = a,
 \label{integration2}
\end{equation}
where we may identify $a$ as a lattice constant. 
In the continuum integration we have the following relations:
\begin{equation}
 \int_{f(x)}^{f(x+a)} df = \int_x^{x+a} \frac{df}{dx}dx = f \Bigr|_x^{x+a} 
 \simeq a\frac{df}{dx}.
\end{equation}
In comparison with the above continuum relation we can show the 
following exact relation on the oriented lattice:
\begin{equation}
1)\ \  \int_x^{x+a} F(x') dx' = \int_x^{x+a} F\theta = a F(x),
\end{equation}
by using solely the Leibniz rule (\ref{eq:d_def1}) and the 
noncommutativity (\ref{noncommutativity}). 
It is possible to generalize into $D$ dimensional total space:
\begin{equation}
1)\ \  \int F \,dx^1\ldots dx^D = \int F \theta^1 \ldots \theta^D
    = \sum_x F_x,
\end{equation}
where we take $a=1$ in the general case. 

The integration on the symmetric lattice can be defined in a similar way. 
The relation (\ref{integration2}) on the symmetric lattice is 
\begin{equation}
2)\ \ \int_x^{x+a} dx^\mu
   = a\int_x^{x+a}(\theta^{\mu} - \theta^{-\mu})
   = x^\mu\Big|_x^{x+a} = a, 
\end{equation}
which naturally leads the definition of $1$-dimensional integral 
(along $\mu$ direction)
\begin{eqnarray}
 2)\ \ \int_x^{x+a} \theta^\mu = \alpha,
   \quad \int_x^{x+a} \theta^{-\mu} = (\alpha-1).
\end{eqnarray}
Here $\alpha$ is an arbitrary parameter with $0\leq \alpha \leq 1$.
Integrals of general $1$-forms in $1$ dimension are
\begin{eqnarray}
 2)\ \ \int_x^{x+a} f \theta^\mu &=&  \alpha f(x), \label{eq:s-int-p} \\
 \int_x^{x+a} f \theta^{-\mu} &=& (\alpha-1)f(x+a)\label{eq:s-int-m}
\end{eqnarray}
which are consistent with the relation (\ref{integration1}).
Combining (\ref{eq:s-int-p}) and (\ref{eq:s-int-m}), we obtain
\begin{equation}
 2)\ \ \int_x^{x+a} f \theta^{\epsilon\mu}
  = \left( \alpha - \frac{1-\epsilon}{2}\right) f(x + \frac{1-\epsilon}{2}a).
\end{equation}

To extend into higher dimensional integral, we impose
\begin{equation}
2)\ \  \int \theta^\mu \theta^{-\mu} \equiv 0,
\label{null-area}
\end{equation}
which has geometrical interpretation. 
We may identify $\theta^{\mu}\theta^{\nu}$ as an area element spanned 
by the differential $\theta^{\mu}$ and $\theta^{\nu}$. 
Then the area element $\theta^{\mu}\theta^{-\mu}$ which is 
spanned by $\theta^{\mu}$ and $\theta^{-\mu}$ does not span an area but 
a line element and thus should vanish. 

\begin{figure}
 \begin{minipage}{.5\linewidth}
 \includegraphics[width=\linewidth]{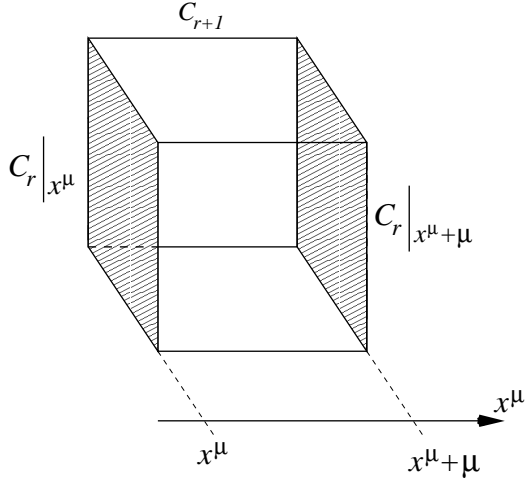}  
 \end{minipage}
 \begin{minipage}{.4\linewidth}
   \caption{r-cubes on $x^\mu$, $C_r|_{x^\mu}$, and on $x^\mu + \mu$, 
$C_r|_{x^\mu+\mu}$, compose (r+1)-cube.}
 \label{r+1-cube}
 \end{minipage}
\end{figure}

In $r$-dimensional hyper cube, we define the integral as
\begin{eqnarray}
 && 2)\ \ \lefteqn{\int_{C_r} f \theta^{\epsilon_1 \mu_1}
  \ldots \theta^{\epsilon_r \mu_r}}\nn\\ 
  &=& \left( \alpha - \frac{1-\epsilon_1}{2}\right)
      \ldots \left( \alpha - \frac{1-\epsilon_r}{2}\right)
      f( x + \frac{1-\epsilon_1}{2}\mu_1 + \cdots
           + \frac{1-\epsilon_r}{2}\mu_r )
\end{eqnarray}
with $\mu_i \neq \mu_j$. 
The $r$-cube is defined as
\begin{eqnarray}
 C_r = \{ x'| x^{\mu_i} \leq x'^{\mu_i} \leq x^{\mu_i} + a, i=1,\ldots,r\},
  \label{eq:r-cube}
\end{eqnarray}
which is pictorially shown in Fig. 
\ref{r+1-cube}. 
The integral vanishes unless a space spanned by
$\theta^{\epsilon_i\mu_i}$'s are identical to that of $C_r$.
In order to check that this definition of the integral on the lattice 
is consistent we have tried to prove Stokes theorem. 
The proof is given in Appendix A.

We next define the inner product of differential forms on the lattice. 
For the oriented lattice 1) inner product of differential forms can be defined 
in a standard way with a slight modification due to the noncommutativity. 
We define the Hodge dual of $p$-form in $D$ dimension as follows:
\begin{eqnarray}
 1)\ \  * \bigl( f_{\mu_1 \ldots \mu_p}^{(p)} 
   \theta^{\mu_1} \ldots \theta^{\mu_p} \bigr)
    = \bigl( T_{-\mu_1 \cdots -\mu_p} f_{\mu_1 \ldots \mu_p}^{(p)} \bigr)
       *  \theta^{\mu_1} \ldots \theta^{\mu_p} \nonumber \\
    = \bigl( T_{-\mu_1 \cdots -\mu_p} f_{\mu_1 \ldots \mu_p}^{(p)} \bigr)
    \frac{1}{(D-p)!} \varepsilon^{\mu_1\ldots\mu_p}{}_{\nu_1\ldots\nu_{D-p}}
	 \theta^{\nu_1} \ldots \theta^{\nu_{D-p}}.
	 \label{inner-product1}
\end{eqnarray} 
The inner product of inhomogeneous differential form $f$ and $g$ can be given 
by 
\begin{eqnarray}
 1)\ \  (f,g)\ \  = \ \ \int f^* *g\Bigr|_\Dform \ \ &=& \ \ 
   \int \sum_{p=0}^D \frac{1}{p!} 
   f_{\mu_1 \ldots \mu_p}^{(p)*}(g^{(p)})^{\mu_1 \ldots \mu_p} 
   \theta^1 \ldots \theta^D \nn \\
   \ \ &=& \ \ \sum_x\sum_{p=0}^D \frac{1}{p!} 
   f_{x,\mu_1 \ldots \mu_p}^{(p)*}(g^{(p)}_x)^{\mu_1 \ldots \mu_p},
\end{eqnarray}
where 
\begin{eqnarray}
 1)\ \ \ \  f\ \  &=&  \ \ \sum_{p=0}^D \frac{1}{p!} 
   f_{\mu_1 \ldots \mu_p}^{(p)}\theta^{\mu_1} \ldots \theta^{\mu_p}, \nn \\
   g\ \  &=&  \ \ \sum_{p=0}^D \frac{1}{p!} 
   g_{\mu_1 \ldots \mu_p}^{(p)}\theta^{\mu_1} \ldots \theta^{\mu_p},
\end{eqnarray}
with 
\begin{eqnarray}
f_{\mu_1 \ldots \mu_p}^{(p)} = 
\sum_x f_{x,\mu_1 \ldots \mu_p}^{(p)} \e^x.
\end{eqnarray}
The shift operator $T_{-\mu_1 \cdots -\mu_p}$ in the definition of 
Hodge dual (\ref{inner-product1}) is necessary to reproduce the standard 
form of the inner product for differential forms. 
To be more explicit the inner product for the bases of the differential 
form can be given by 
\begin{eqnarray}
1)\ \ \ \   ( \theta^{\mu_1}\ldots \theta^{\mu_p} ,
          \theta^{\nu_1}\ldots \theta^{\nu_q})
  = \sum_x \delta_{pq} \delta^{\mu_1}_{[\nu_1}\ldots \delta^{\mu_p}_{\nu_p]},
  \label{base-inner-product1}
\end{eqnarray}
where $[\nu_1 \ldots \nu_p]$ denotes an antisymmetric sum of the 
suffices.

In the case of symmetric lattice inner product should be defined in 
a different way from the case of oriented lattice since there are $2D$ 
differential 1-forms in $D$ dimension. 
We impose the following conditions: 
\begin{eqnarray}
2)\ \ \ (dx^\mu,dx^\nu) &=& 
(\tau^\mu,\tau^\nu) \ \  = \ \  \sum_x\delta^{\mu\nu} , \nonumber 
\\ 
(dx^\mu,\tau^\nu) &=& (\tau^\mu,dx^\nu) = 0 ,
\end{eqnarray}
where $dx^\mu=\theta^\mu-\theta^{-\mu}$ and 
$\tau^\mu=\theta^\mu+\theta^{-\mu}$ 
as defined in eqs. (\ref{dx-of-2)}) and (\ref{tau-of-2)}), respectively. 
In the representation space of this inner product $dx^\mu$ behaves as 
real while $\tau^\mu$ behaves as pure imaginary and the $\ast$-conjugation 
plays the role of complex conjugate in accordance with the relation 
(\ref{real-imaginary-star-relation}). The space of $dx^\mu$ 
and $\tau^\mu$ are, however, orthogonal. 
The above conditions naturally leads the following relations:
\begin{eqnarray}
2)\ \ \ (\theta^\mu,\theta^\nu) &=& (\theta^{-\mu},\theta^{-\nu}) = 
 \frac{1}{2} \sum_x \delta^{\mu\nu},\nn \\
(\theta^\mu,\theta^{-\nu}) &=& (\theta^{-\mu},\theta^\nu) = 0.
\end{eqnarray} 
Then the symmetric lattice counterpart of the formula 
(\ref{base-inner-product1}) can be obtained by extending the above 1-form 
inner product formulae to higher forms, 
\begin{eqnarray}
 2)\ \ \bigl(\theta^{\epsilon_1 \mu_1} \ldots \theta^{\epsilon_p \mu_p}  ,
          \theta^{\epsilon'_1 \nu_1} \ldots \theta^{\epsilon'_q \mu_q} \bigr)
 = \sum_x  \frac{1}{2^p} \delta_{pq}
     \delta^{\epsilon_1\mu_1}_{[\epsilon'_1\nu_1} \ldots
            \delta^{\epsilon_p\mu_p}_{\epsilon'_p \nu_p]} .
 \label{eq:inner-p-Dimakis}
\end{eqnarray}
It should be noted that the naive Hodge dual operation on the differential 
forms does not make sense since we have $2D$ 1-forms for the symmetric 
lattice.  


\section{Dirac-K\"ahler Fermion and Staggered Fermion}

\setcounter{equation}{0}

Here we summarize the well-known results of Dirac-K\"ahler fermion 
formulation formulated by Clifford product and staggered fermion 
formulation on the lattice\cite{Becher:1982ud,Banks:1982iq,Rabin:1982qj,
Benn:1983sr,Bullinaria:1986ed}\cite{Kogut:1975ag,Susskind:1977jm}
\cite{Kawamoto:1981hw,Kluberg-Stern:1983dg,Gliozzi:1982ib}. 
Dirac-K\"ahler fermion incorporates inhomogeneous differential forms 
to describe Dirac fermions. Here we consider the spacetime is even 
dimensional Euclidean space.
Gamma matrices satisfies the Clifford algebra:
\begin{equation}
\{\gamma^\mu, \gamma^\nu \}=2\delta^{\mu\nu}, 
\end{equation}
where $(\gamma^\mu)^\dagger = \gamma^\mu$ and 
$\gamma_5 = \gamma^1 \ldots \gamma^D$, where $D$ is the spacetime dimension. 
We denote the unit vector of $\mu$-direction simply by $\mu$.

We first note the following well known relations on the flat continuous 
space:
\begin{equation}
 (d-\delta)^2 = \partial^\mu\partial_\mu = (\gamma^\mu\partial_\mu)^2,
\end{equation}
where $\delta$ is the adjoint of the exterior derivative $d$ 
and has the following form in even dimensions:
\begin{eqnarray}
\delta = -*d* \equiv -e^\mu \partial_\mu.
\end{eqnarray}
with $*$ as Hodge star dual.
Here $e_\mu=\partial_L/\partial_L(dx^\mu)$ is a left derivative differential 
operator of the differential $dx^\mu$ and thus contracts vector suffices, 
in particular $e^\mu dx^\nu = \delta^{\mu\nu}$.
The above relation suggests the
following correspondences: 
\begin{eqnarray}
d ~-~\delta~=~\partial_\mu(dx^\mu\wedge ~+~e^\mu) ~&\sim&~
\gamma^\mu\partial_\mu,  \nonumber \\
dx^\mu \vee \equiv dx^\mu\wedge ~+~e^\mu~&\sim&~\gamma^\mu, 
\label{dxClifford}
\end{eqnarray}
where we have introduced the simplest version of Clifford product 
$\vee$.
We can then show that the Clifford product satisfies the Clifford
algebra: 
\begin{eqnarray}
\{ dx^\mu,dx^\nu\}_{\vee}~ \equiv dx^\mu \vee dx^\nu + 
dx^\nu \vee dx^\mu= ~ 2 \delta^{\mu\nu}.
\end{eqnarray}

The Clifford product $\vee$ for arbitrary inhomogeneous forms
$\Phi, \Psi$ is defined as follows\cite{Kahler:1962}:
\begin{eqnarray}
 \Phi \vee \Psi
   &\equiv& \sum_{p=0}^D \frac{1}{p!} (-)^{p(p-1)/2}\, \{ \eta^p
          \left( e_{\mu_1} \ldots e_{\mu_p} \Phi \right)\} \wedge
	  \left( e^{\mu_1} \ldots e^{\mu_p} \Psi \right) \nn\\
   &=& \sum_P (-)^{p(p-1)/2}\, \{\eta^p (e_{\!P}^\ssstrut \Phi)\}
             \wedge (e^P \Psi),
    \label{eq:Clifford-normal}
\end{eqnarray}
where we have introduced shorthand notations 
$\sum_P=\sum_{p=0}^D \frac{1}{p!}$ and 
$e^{P}=e^{\mu_1} \ldots e^{\mu_p}$. 
The sign operator $\eta$ is defined as 
\begin{equation}
\eta(\mbox{$p$-form})=(-)^p(\mbox{$p$-form}).
\label{sign-factor-eta}
\end{equation} 
The sign operator and the sign factor are necessary for the associativity 
of the Clifford products.

It is convenient to introduce the following basis:
\begin{eqnarray}
  Z^i{}_{\!(j)}
   &=& \sum_{p=0}^D \frac{1}{p!}
        \left( \gamma_{\mu_1}{}^T \gamma_{\mu_2}{}^T
	         \ldots \gamma_{\mu_p}{}^T \right)^i{}_{\!(j)}\,\,
	dx^{\mu_1} \wedge dx^{\mu_2} \wedge \cdots \wedge dx^{\mu_p} \nn\\
   &=& \sum_P \left( \gamma^T \right)_{\!P}\! {}^i{}_{\!(j)}\,\, dx^P,
    \label{eq:Zdef}
\end{eqnarray}
where we have introduced the shorthand notation of 
$\left( \gamma^T \right)_{\!P}$ and $dx^P$.
The transpose of gamma matrices is introduced for the later convenience. 
This basis represents the Clifford product of differential forms as 
products of gamma matrices 
\begin{equation}
 dx^\mu \vee Z^i{}_{\!(j)} = (\gamma^\mu){}_k{}^i{} Z^k{}_{\!(j)},
  \qquad Z^i{}_{\!(j)} \vee dx^\mu = Z^i{}_{\!(l)} 
  (\gamma^\mu){}_{(j)}{}^{(l)}.
\end{equation}
Dirac-K\"ahler equation for a inhomogeneous differential form 
$\Psi = \psi_i{}^{(j)} Z^i{}_{\!(j)}$ is readily obtained by the 
Clifford product: 
\begin{equation}
 (d + m )\vee \Psi
  = \left[ \gamma^\mu{}_i{}^k \del_\mu \psi_k{}^{(j)} + m \psi_i^{(j)}
      \right] Z^i{}_{\!(j)}= 0,
  \label{eq:DiracKahler}
\end{equation}
where the constant $0$-form mass term $m$ is introduced. 
This equation describes $2^{D/2}$ Dirac fermions labeled by $(j)$ which
are commonly identified as ``flavor'' suffix. 
The Dirac-K\"ahler fermion formulation turns the antisymmetric tensor 
fields with differential forms into Dirac spinor fields with flavor copies.  
There is an interesting formulation where the ``flavor'' suffix can be 
identified as the extended supersymmetry suffix in the twisted version of 
generalized topological gauge theory. 
It was pointed out that the twist of the topological gauge theory 
is nothing but the Dirac-K\"ahler fermion 
formulation\cite{Kawamoto:1999zn,K-K-U}.


We need to introduce an inner product to generate an action with 
Dirac-K\"ahler fermion.
A positive definite inner product for 
$\Phi  = \sum_p \frac{1}{p!}\phi_{\mu_1\ldots\mu_p} dx^{\mu_1} \ldots
 dx^{\mu_p}$ 
and $\Psi  = \sum_p \frac{1}{p!}\psi_{\mu_1\ldots\mu_p} dx^{\mu_1} \ldots
 dx^{\mu_p}$ 
is 
\begin{eqnarray}
 (\Phi, \Psi)
   &=& \int {\Phi}^* * \Psi \Bigr|_\Dform \nn \\
   &=& \int (\zeta{\Phi}^* \vee \Psi) *1
         = \int (\zeta{\Phi}^* \vee \Psi)\Bigr|_{\zform} *1 \nn \\
   &=& \int \sum_p \frac{1}{p!} {\phi}^*_{\mu_1\ldots\mu_p}
        \psi^{\mu_1\ldots\mu_p} *1, 
\end{eqnarray}
where $*$ is Hodge star operation and thus $*1$ is a volume form. 
We have introduced another sign operator $\zeta$ which acts as
\begin{equation}
\zeta (p\mbox{-form}) =(-)^{p(p-1)/2}(p\mbox{-form}).
\label{sign-factor-zeta}
\end{equation} 

This definition of the inner product naturally leads $(\Phi^*, \Phi) \geq 0$.
For the basis $Z^i{}_{\!(j)}$ we have the following orthogonality relation:
\begin{equation}
 (Z^k
 {}_{(l)}, Z^i{}_{\!(j)}) = 2^{D/2} \delta^{(l)}_{(j)} 
 \delta^i_k \int *1,
\end{equation}
where we have used the completeness of gamma matrices, 
\begin{equation}
  \sum_{p=0}^D \sum_{\{\mu_i\}}\frac{1}{p!} (-)^{p(p-1)/2}
     (\gamma_{\mu_1}^T \ldots \gamma_{\mu_p}^T)^{(l)}{}_k
     (\gamma_{\mu_1}^T \ldots \gamma_{\mu_p}^T)^i{}_{\!(j)}
   = 2^{D/2} \delta^{(l)}_{(j)} \delta^i_k.
\end{equation}

Defining $\Psibar = \psibar_{(j)}{}^i Z^{(j)}{}_i$, we obtain 
the free fermion action of Dirac-K\"ahler fermion 
\begin{eqnarray}
 S 
   &=& \int \Psibar \vee (d+m) \vee \Psi \Bigr|_\zform *1\\
   &=& 2^{D/2} \int \sum_{(j)} \psibar_{(j)}
       ( \gamma^\mu \del_\mu + m) \psi^{(j)} *1.
    \label{eq:normal-D-K-action}
\end{eqnarray}

We next consider $D$ dimensional Euclidean square lattice and 
derive staggered fermion action from naive fermion action on the lattice. 
These are well-known derivations from early time of lattice gauge 
theory\cite{Kawamoto:1981hw,Kluberg-Stern:1983dg,Gliozzi:1982ib}.
If we naively discritize the Dirac equation on the square lattice 
we obtain
\begin{equation}  
S_F=\frac{1}{2}\sum_{x,{\mu}}[\overline{\psi}(x)\gamma_\mu\psi
    (x+{\mu})-\overline{\psi}(x+{\mu})\gamma_\mu\psi(x)],
\label{eqn:naive-ferm}
\end{equation}
where $x$ denotes the lattice site with integer coordinates 
$x_1,x_2,\cdots,x_D$ in units of lattice spacing which is taken to be 1.

One of the authors and Smit proposed the following transformation which 
relates the naive fermion formulation and Kogut-Susskind fermion 
formulation\cite{Kawamoto:1981hw}: 
\begin{equation}  
\psi(x)=A(x)\chi(x),~~~~~~~~~~~\overline{\psi}(x)=
\overline{\chi}(x)A^\dagger(x),
\label{eqn:KS-trans}      
\end{equation}
where 
\begin{equation}
A(x)=\gamma_1^{x_1}\gamma_2^{x_2}\cdots\gamma_d^{x_D}.
\label{eqn:KS-mat}
\end{equation}
Then the naive fermion formulation of Dirac action leads to
\begin{equation}  
S_F=\frac{1}{2}\sum_{x,{\mu},(j)}\eta_\mu(x)
[\overline{\chi}^{(j)}(x)\chi^{(j)}(x+{\mu})-
\overline{\chi}^{(j)}(x+{\mu})\chi^{(j)}(x)], 
\label{eqn:stag-ferm}
\end{equation}
with 
\begin{equation}
\eta_\mu(x)=(-1)^{x_1+x_2+\cdots+x_{\mu-1}}.
\label{eqn:KS-mat}
\end{equation}
This formulation is commonly called staggered fermion formulation of 
lattice QCD. The $\gamma$ matrices in the naive fermion formulation have now 
disappeared in the staggered fermion formulation. 
Instead we get $2^{[D/2]}$ copies of spinors.

We now consider $D$-dimensional hypercube defined on the lattice, 
with its origin at the site $2y$ and corners 
\begin{equation}
    x_\mu=2y_\mu + h_\mu, ~~~~~h_\mu=0~ \hbox{or}~ 1,~~~
    \mu=1,\cdots,D.\nonumber 
\label{eqn:double-size}
\end{equation}
Thus $h$ labels the set of $2^D$ $D$-dimensional vectors which 
point on the corners of the hypercube. 
Then $\eta_\mu(x) = \eta_\mu(H)$ with $H=\{h_1,h_2, \cdots h_D\}$.
We introduce the following notations for the convenience:
\begin{eqnarray} 
\chi(2y+h)&\equiv& \chi_H(y), \nonumber \\
\overline\chi(2y+h)&\equiv& \overline\chi_H(y). 
\end{eqnarray} 
Here we do not consider the suffix $(j)$ dependence anymore. 
Then $S_F$ can be written in the following 
form\cite{Kluberg-Stern:1983dg,Gliozzi:1982ib}:
\begin{eqnarray}  
S_F=\sum_{y,\mu,H}\eta_\mu(H)\overline{\chi}_H(y) \{
  \delta_{h_\mu,0}(\chi_{H+{\mu}}(y) &-& 
  \chi_{H+{\mu}}(y-{\mu}))\nn\\
 + \delta_{h_\mu,1}(\chi_{H-{\mu}}(y+{\mu}) &-& 
 \chi_{H-{\mu}}(y)) \},
\label{eqn:double-space-action}
\end{eqnarray}
where $H\pm{\mu} = (h_1, \cdots, h_\mu\pm 1, \cdots, h_D)$.
We then introduce the following redefinition of fermion fields:
\begin{eqnarray}
\chi_H(y) &=& \{(\gamma_1^T)^{h_1} (\gamma_2^T)^{h_2}\cdots 
(\gamma_D^T)^{h_D}\}_{ij}\psi_{ij}(y) \nn\\
\overline{\chi}_H(y) &=& \{(\gamma_1^T)^{h_1} (\gamma_2^T)^{h_2}\cdots 
(\gamma_D^T)^{h_D}\}_{ij}^\dagger \overline{\psi}_{ij}(y).
\end{eqnarray}
We can then obtain the final form of the staggered fermion action which 
includes free Dirac fermion action:
\begin{eqnarray}
 S_F
 = 2^{D/2-1}\sum_y \sum_\mu \tr \Bigl\{
      \psibar(y) \gamma_\mu
         \nabla_\mu\psi(y)
      {}+ \psibar_{(j)}(y) \gamma_5^\dagger
         \Delta_\mu\psi(y)
          \gamma_5 \gamma_\mu \Bigr\},
\label{eq:staggered-action-original}
\end{eqnarray}
where 
\begin{eqnarray}
\nabla_\mu \psi(y) &=& \frac{1}{2}(\psi(y+{\mu})-\psi(y-{\mu}))\nn\\
\Delta_\mu \psi(y)&=& \frac{1}{2}(\psi(y+{\mu})-2\psi(y)+
\psi(y-{\mu})).
\label{def-difference-operator}
\end{eqnarray}
This staggered fermion action includes $2^{D/2}$ copies of Dirac fermions. 
The Dirac fermion part of this action can be interpreted as Kogut-Susskind 
fermion action while this action includes the second derivative counterpart 
as well. The second derivative part is higher order 
than the Dirac fermion part with respect to the lattice constant.


\section{Dirac-K\"ahler Fermion from Clifford Product on the Oriented Lattice}

\setcounter{equation}{0}

In order to formulate Dirac-K\"ahler fermion on a lattice 
we need to introduce the differential form and Clifford product on 
the lattice. 
Because of the noncommutativity a naively extended version of Clifford
product of the continuum formulation to the lattice 
spoils the associativity. 
In this section we construct an associative Clifford product on the 
oriented lattice and generate Dirac-K\"ahler fermion. 
We need to mention that the same Clifford algebra and the corresponding 
1-form Clifford product of this section were given by Vaz\cite{Vaz1,Vaz2}. 
Here we explicitly give the definition of the Clifford product for 
general higher forms. 
We find that the Hermiticity and the associativity of the Clifford 
product are not compatible in formulating Dirac-K\"ahler fermion on 
the oriented lattice.

In order to keep the associativity of the Clifford product, we have
to carefully compensate the noncommutative effect by inserting some
shift operators. 
As we can see in the definition of the continuum Clifford product 
(\ref{eq:Clifford-normal}) for a pair of general forms, 
we find necessity to keep the same noncommutative nature for each terms 
after the operation of $e_\mu$ contractions. 
To be consistent with the nature of the contraction operator 
$e_\mu dx^\nu=\delta_\mu^\nu$ and the noncommutative nature 
between $dx^\mu=\theta^\mu$ and a function $f_x$ on the oriented lattice:
\begin{equation}
 dx^{\mu} f_x = T_\mu f_x dx^{\mu} = f_{x+\mu} dx^{\mu}, 
\end{equation}
we impose the following noncommutativity on the contraction operator:
\begin{equation}
 e^{\mu} f_x = T_{-\mu}f_x e^{\mu} = f_{x-\mu} e^{\mu}. 
 \label{contraction-operator1}
\end{equation} 
The continuum Clifford product (\ref{dxClifford}) should now be modified 
on the oriented lattice:
\begin{equation}
1) \ \ \ dx^\mu \vee{} = dx^\mu \wedge{} + \Tleft{-\mu} e^\mu \Tright{\mu},
\label{oriented-clifford}
\end{equation}
where the last term has the same noncommutative nature as the first term. 
The shift operators with right(left) arrow act all the $x$ dependent 
functions located in the forward(backward) direction:
\begin{equation}
 f \Tright{\mu}g = f (T_\mu g) \Tright{\mu}
 \qquad f \Tleft{\mu} g = \Tleft{\mu} (T_\mu f) g.
 \label{shiftoperators}
\end{equation}
It should, however, be noted that the range of the operation of the 
shift operators $\Tright{\mu}$ and $\Tleft{\mu}$ should be 
terminated when the range of the operation comes outside of 
an integration. 
The similar definition of this 1-form Clifford product without 
$\Tleft{-\mu}$ in (\ref{oriented-clifford}) was given by 
Vaz\cite{Vaz1,Vaz2}. We, however, point out at this stage that 
the necessity of $\Tleft{-\mu}$ is crucial in the proof of 
associativity of Clifford product for general higher forms on the 
oriented lattice, which will be given below. 
The oriented lattice version of 1-from Clifford algebra is given by 
\begin{eqnarray}
\{ dx^\mu,dx^\nu\}_{\vee}~ = ~ \Tleft{-\mu} 2 \delta^{\mu\nu} \Tright{\mu}, 
\end{eqnarray}
where the shift operators in the righthand side are necessary to 
compensate the noncommutative nature of the 2-form on the left 
hand side. 

Generalizing the above treatment to define Clifford product on the 
oriented lattice, 
we claim that the following Clifford product which has shift operators 
in addition to the continuum version (\ref{eq:Clifford-normal}) 
satisfies associativity:
\begin{eqnarray}
1) \ \ \ \lefteqn{(f \,dx^K) \vee (g \,dx^L)}\hspace{1em}\nn\\
  &\equiv& \sum_p \frac{1}{p!} (-)^{p(p-1)/2}
        \Bigl( \eta^p \Tleft{-\mu_1} \ldots \Tleft{-\mu_p}
               e_{\mu_1} \ldots e_{\mu_p} f\,dx^K \Bigr)
       \nonumber\\[-.8em]
  && \hspace{12em}
      {} \wedge \Bigl( g e^{\mu_1} \ldots e^{\mu_p} dx^L
                    \Tright{\mu_1} \ldots \Tright{\mu_p} \Bigr)
     \label{eq:oriented-Clifforddef}\nn\\
  &=& \sum_P (-)^{p(p-1)/2}
        \Bigl( \eta^p \Tleft{-P} e_P f\, dx^K \Bigr)
	{}\wedge \Bigl( g e^P dx^L \Tright{P} \Bigr),
\end{eqnarray}
where we have introduced the shorthand notations 
$e_H$, $dx^K$, $\Tleft{-H}$ and $\Tright{H}$.
The coefficient functions $f, g$ are in general anti-symmetric 
tensor fields and may have tensor suffices $\mu_1 \ldots \mu_h$. 
We, however, omit the suffix dependence for simplicity.

The most important property of this Clifford product is
\begin{equation}
 (f\, dx^K \vee g\,dx^L) h = (T_K T_L h) (f\, dx^K \vee g\,dx^L),
  \label{eq:oriented-Clifford}
\end{equation}
which shows that all the inhomogeneous terms in the Clifford product 
of (\ref{eq:oriented-Clifforddef}) have the same noncommutative nature. 
Without this property the Clifford product could not be associative.

From (\ref{eq:oriented-Clifford}) and $ (dx^K f) \vee dx^L = dx^K \vee
(f\,dx^L)$, we obtain
\begin{eqnarray}
 (f\, dx^K \vee g\,dx^L) \vee h\,dx^M
   &=& f (T_K g )(T_K T_L h) \bigl\{ (dx^K \vee dx^L)\vee dx^M \bigr\}, 
   \nonumber \\
 f\, dx^K \vee ( g\,dx^L \vee h\,dx^M)
   &=& f (T_K g )(T_K T_L h) \bigl\{ dx^K \vee (dx^L \vee dx^M) \bigr\}.
\label{eq:oriented-associativity0}
\end{eqnarray}
In order to prove the associativity of the Clifford product defined on 
the oriented lattice we simply need to prove
\begin{equation}
 (dx^K \vee dx^M) \vee dx^L = dx^K \vee (dx^M \vee dx^L).
  \label{eq:oriented-associativity}
\end{equation}
We give the proof of (\ref{eq:oriented-associativity}) in appendix 
B.
Due to the presence of the shift operators $\Tleft{-\mu}$ and $\Tright{\mu}$, 
the proof of the associativity is not trivial.

%

Having defined the oriented lattice version of the Clifford product, 
we define Dirac operator as
\begin{equation}
 d\vee{} = \sum_\mu dx^\mu \vee \del_{-\mu}.
\end{equation}
The square of this Dirac operator leads
\begin{equation}
 d\vee d\vee{}
  = \sum_\mu\Tleft{-\mu} \Bigl( \Tright{\mu} + \Tright{-\mu} -2 \Bigr)
  = \sum_\mu \Tleft{-\mu} \Bigl( T_\mu + T_{-\mu} -2 \Bigr),
\end{equation}
which is similar to the usual lattice Laplacian except for 
the extra shift operator $\Tleft{-\mu}$.

Here we introduce the same bases $Z^i{}_{\!(j)}$ as the continuum version 
(\ref{eq:Zdef}) and expand fermionic fields, which are composed of 
fermionic differential forms, by these bases:
\begin{eqnarray}
 Z^i{}_{\!(j)}
  &=& \sum_P \left( \gamma^T \right)_{\!P}\! {}^i{}_{\!(j)}\,\, dx^P \\
 \Psi &=& \psi_i{}^{(j)} Z^i{}_{\!(j)} 
 = \psi_i{}^{(j)}({\tilde{Z}^i_{\mu (j)}} + 
 dx^\mu(\gamma_\mu^T{\tilde{Z}_\mu})^i{}_{(j)}),
\end{eqnarray}
where $\tilde{Z}^i_{\mu (j)}$ is the similar basis as (\ref{eq:Zdef}) 
except that it includes neither $dx^\mu$ nor $\gamma^\mu$.
We introduce the last expression for the later convenience. 
It should be noted that 
$\psi_i{}^{(j)} Z^i{}_{\!(j)} \neq  Z^i{}_{\!(j)} \psi_i{}^{(j)}$ 
due to the noncommutativity.

We then obtain 
\begin{eqnarray}
 \lefteqn{
   d \vee \Psi
  }\nonumber\\
  &=& \sum_\mu dx^\mu \vee ( 1- T_{-\mu})\Psi \nn\\
  &=& \frac{1}{2}\sum_\mu (\gamma_\mu)_i{}^k
       \Bigl[ (T_\mu -1)\psi_k{}^{(j)}
              + \Tleft{-\mu}(1-T_{-\mu})\psi_k{}^{(j)}\Tright{\mu} \Bigr]
       Z^i{}_{\!(j)} \nn\\*
  &&  
      {} + \frac{1}{2}\sum_\mu (\gamma_5^\dagger)_i{}^k
       \Bigl[ (T_\mu -1)\psi_k{}^{(l)}
              - \Tleft{-\mu}(1-T_{-\mu})\psi_k{}^{(l)}\Tright{\mu} \Bigr]
       (\gamma_5 \gamma_\mu)_{(l)}{}^{\!(j)}  Z^i{}_{\!(j)},
       \qquad\quad
       \label{eq:nc.dPsi}
\end{eqnarray}
where we have used the following relations:
\begin{eqnarray}
dx^\mu {\tilde{Z}^i_{\mu (j)}} + (\gamma_\mu^T)^i{}_{k} 
{\tilde{Z}^k_{\mu (j)}}
&=& (\gamma_\mu^T)^i{}_{k} Z^k{}_{(j)} \nn \\
dx^\mu {\tilde{Z}^i_{\mu (j)}} - (\gamma_\mu^T)^i{}_{k} 
{\tilde{Z}^k_{\mu(j)}}
&=& (\gamma_5^T)^{\dagger i}{}_{l} Z^l{}_{(k)} 
(\gamma_\mu^T\gamma_5^T)^{(k)}{}_{(j)}.
\label{z_mu-relation}
\end{eqnarray}

In order to obtain Dirac-K\"ahler fermion action we need to establish  
the relation between the inner product and the Clifford product. 
Identifying $dx^\mu=\theta^\mu$ on the oriented lattice, 
we find 
\begin{eqnarray}
\int \zeta (dx^{\mu_1}\wedge \cdots \wedge dx^{\mu_p}) &\vee& 
(dx^{\nu_1}\wedge \cdots \wedge dx^{\nu_q}) *1 \nn \\
\ &=& \ (dx^{\mu_1} \cdots dx^{\mu_p},dx^{\nu_1} \cdots dx^{\nu_q}) \nn \\
\ &=& \ \sum_x \delta_{pq} \delta^{\mu_1}_{[\nu_1}\ldots 
\delta^{\mu_p}_{\nu_p]},
\end{eqnarray} 
where we have already introduced the inner product in 
(\ref{base-inner-product1}).

We introduce an action with the Clifford product 
\begin{equation}
 S = \int \bigl( \Psibar \vee (d+m) \vee \Psi \bigr) *1,
\end{equation}
where $\Psi = \psi_i{}^{(j)} Z^i{}_{\!(j)}$ and 
$\Psibar = Z^{(j)}\!{}_i\,\psibar_{(j)}{}^i$. 
The mass term satisfies the following obvious relation:
\begin{equation}
 \int \bigl(\Psibar \vee \Psi \bigr)*1
   = 2^{D/2} \sum_x \psibar_{(j)}{}^i(x) \psi_i{}^{(j)}(x).
\end{equation}
The above action then leads
\begin{equation}
 S
  = 2^{D/2} \sum_x \psibar_{(j)}{}^i(x)
       \Bigl[ \sum_\mu (\gamma_\mu)\,{}_i{}^k
              \bigl\{ \psi_k{}^{(j)}(x+\mu) - \psi_k{}^{(j)}(x) \bigr\}
            + m \psi_i{}^{(j)}(x)
       \Bigr].\label{eq:o-ncD-Kaction1}
\end{equation}
Since the difference of the fermion fields is asymmetric, this action 
is not Hermitian and thus has ill-defined propagator. 
Hence naively extended action of the Dirac-K\"ahler fermion formulation 
on the oriented lattice lacks Hermiticity. 
In order to recover the Hermiticity we can introduce another type of 
inner product where Hodge star operation can be defined in such a way 
that Hermiticity is recovered\cite{kanamori1}. 
This type of definition of the inner 
product, however, sacrifices the associativity of the Clifford product.  

\section{Clifford Product and Hermiticity on the Symmetric Lattice}

\setcounter{equation}{0}

In the last section we have pointed out that Hermiticity and associativity 
of the Clifford product cannot be accommodated in a compatible way on the
oriented lattice. Here we propose to define an 
associative Clifford product on the symmetric lattice. 
The Dirac-K\"ahler fermion action can be derived by this Clifford 
product to be compatible with Hermiticity. 
To define a new Clifford product on the symmetric lattice we impose the 
following two conditions; i) associativity of Clifford product and
ii) positivity of inner product, $\zeta \Phi^* \vee
\Phi\bigr|_\zform \geq 0$.

On the oriented lattice there is single $\theta^\mu$, on a link connecting 
$x$ and $x+\mu$, which can be identified as 1-form differential $dx^\mu$.
There are, however, $\theta^\mu$ and $\theta^{-\mu}$ on a link of 
symmetric lattice and the 1-form differential is identified as 
$dx^\mu=\theta^\mu - \theta^{-\mu}$. 
On the symmetric lattice there is an arbitrariness for the definition 
of the contraction operator $e^\mu$ since there correspond two 
1-form differentials $\theta^\mu$ and $\theta^{-\mu}$ on a link. 
We generalize to define 
\begin{equation}
e_\mu = 
\alpha\ddx{\theta^{\epsilon\mu}}\Tright{\epsilon\mu} + \beta
\ddx{\theta^{-\epsilon\mu}} \Tright{-\epsilon\mu},
\end{equation}
where $\epsilon = \pm$ and thus $\epsilon \mu = +\mu$ or 
$-\mu$ and $\alpha$ and $\beta$ are constants to be fixed 
later. 
The shift operators $\Tright{\epsilon\mu}$ and $\Tright{-\epsilon\mu}$ 
act in the same way as (\ref{shiftoperators}).
The contraction operator should have the opposite shifting role 
to the differential as in (\ref{contraction-operator1}) 
\begin{equation}
\ddx{\theta^{\pm\mu}} f = (T_{\mp\mu}f) \ddx{\theta^{\pm\mu}},
\end{equation}
which leads 
\begin{equation}
\ddx{\theta^{\epsilon\mu}}\Tright{\epsilon\mu} f = 
f \ddx{\theta^{\epsilon\mu}}\Tright{\epsilon\mu}.
\label{contraction-operator2} 
\end{equation}
We then define the following Clifford product:
\begin{eqnarray}
 \lefteqn{
 f \theta^K \vee g \theta^L 
 }\hspace{1em}\nonumber\\
  &=& \sum_{p=0}^{2D} (-)^{p(p-1)/2}\sum_{\{\epsilon_i \mu_i\}}
     \left\{ \eta^p \prod_{i=1}^p \left(
        \alpha \ddx{\theta^{\epsilon_i\mu_i}} \Tright{\epsilon_i\mu_i}
        + \beta \ddx{\theta^{-\epsilon_i\mu_i}} \Tright{-\epsilon_i\mu_i}
     \right) f \theta^K \right\} \nonumber\\*
  &&
   \hspace{3em}
   {}\wedge \left\{ \prod_{i=1}^p \left(
        \gamma \ddx{\theta^{\epsilon_i\mu_i}} \Tright{\epsilon_i\mu_i}
        + \delta \ddx{\theta^{-\epsilon_i\mu_i}} \Tright{-\epsilon_i\mu_i}
     \right)g \theta^L\right\} ,
   \label{eq:newClifford}
\end{eqnarray}
where we have introduced the short hand notation $\theta^K =
\theta^{\epsilon_1 \mu_1} \ldots \theta^{\epsilon_k \mu_k}$ and 
$T_K = T_{\epsilon_1 \mu_1}\ldots T_{\epsilon_k \mu_k}$ for the later 
convenience.
Here we introduce different parameters for the second contraction 
operator. In the summation it is natural to sum up $2D$ contractions, 
since we have $2D$ 1-forms in the $D$-dimensional symmetric lattice.

Due to the property (\ref{contraction-operator2}) the contraction 
operators in (\ref{eq:newClifford}) do not generate any shift. 
Thus the shifting property of the Clifford product on the symmetric 
lattice is the same as that of wedge product and hence leads the same 
relations as (\ref{eq:oriented-Clifford}) and (\ref{eq:oriented-associativity0}) 
for the oriented lattice.
In order to prove the associativity of Clifford product on the symmetric 
lattice we simply need to prove the associativity of the following shift 
invariant forms as in the case of oriented lattice:
\begin{equation}
 \theta^K \vee (\theta^M \vee \theta^L)
   = (\theta^K \vee \theta^M) \vee \theta^L.
\label{symmetric-lattice-associativity0 }   
\end{equation}
We give the proof in appendix C.
The parameters $\alpha, \beta, \gamma$, and $\delta$ are not restricted
from the associativity.

We now define an inner product of differential forms $\Phi$ and $\Psi$
\begin{equation}
 (\Phi,\Psi) = \int \zeta \Phi^* \vee \Psi \Bigr|_\zform \vol. 
 \label{inner-product}
\end{equation}
We can then show that the positivity of a self inner product of a 
field $\Phi$
\begin{equation}
 (\Phi,\Phi) \geq 0,
 \label{self-positivity}
\end{equation}
imposes constraints on the parameters introduced in the definition 
of the Clifford product. 
Here the $\vol$ is appropriately chosen volume element for the 
symmetric lattice to pick up the 0-form components in the integrand. 
Since there are $2D$ 1-forms and the relation (\ref{null-area}) is 
imposed on the integration of symmetric lattice, 
the definition of volume element needs further cares. 

It is useful to note that the following factorization property holds 
for the Clifford product on the lattice:
\begin{eqnarray}
 \lefteqn{
 \left\{ (\theta^{\epsilon\mu})^a (\theta^{-\epsilon\mu})^b \theta^K \right\}^*
 \vee (\theta^{\epsilon\mu})^a (\theta^{-\epsilon\mu})^b \theta^K
 }\hspace{1em} \nonumber\\
 &=& (-)^{k(a+b)} \left\{ 
    (\theta^{\epsilon\mu})^a{}^* (\theta^{-\epsilon\mu})^b{}^*
    \vee (\theta^{\epsilon\mu})^a (\theta^{-\epsilon\mu})^b  \right\}
    \left( \theta^K{}^* \vee \theta^K \right),
\label{factorization1}    
\end{eqnarray}
where $\theta^K$ contains neither $\theta^{\epsilon\mu}$ nor
$\theta^{-\epsilon\mu}$. 
The proof of this formula is given in appendix C while similar factorization 
property of Clifford product is satisfied even on the oriented lattice as 
shown in (\ref{eq:app-factorization1}). 
Let's consider the case where 
$\Phi=f (\theta^{\epsilon\mu})^a (\theta^{-\epsilon\mu})^b \theta^K$ which 
is general enough for the proof of associativity. 
We can then show 
\begin{eqnarray}
 \lefteqn{
 \left[ \zeta \left\{
      f (\theta^{\epsilon\mu})^a (\theta^{-\epsilon\mu})^b \theta^K
      \right\}^* \right]
 \vee f (\theta^{\epsilon\mu})^a (\theta^{-\epsilon\mu})^b \theta^K
 }\hspace{1em} \nonumber\\
 &=& \left[ \zeta \left\{ T_{K^*}f
      (\theta^{\epsilon\mu})^a (\theta^{-\epsilon\mu})^b \right\}^*
      \vee T_{K^*}f (\theta^{\epsilon\mu})^a (\theta^{-\epsilon\mu})^b \right]
      \left( \zeta \theta^K{}^* \vee \theta^K \right),
 \label{eq:posi-full}
\end{eqnarray}
where we have used the following relations:
\begin{eqnarray}
\zeta(\omega_p\omega_q) &=& (-)^{pq}(\zeta\omega_p)(\zeta\omega_q)\\
(\omega_p\omega_q)^* &=& (-)^{pq}\omega_q^*\omega_p^*,
\end{eqnarray}
with $\omega_p$ and $\omega_q$ as $p$-form and $q$-form, respectively. 
This relation shows that we can confirm the positivity of a general 
form if we can prove the positivity of the term which includes 
only $\theta^{\pm\mu}$ since the Clifford product of 
the general forms can be factorized into the product of these terms. 

For an arbitrary form which includes $\theta^{\pm\mu}$ 
\begin{eqnarray}
F=f_0 + f_+ \theta^\mu + f_- \theta^{-\mu} + \tilde{f} 
\theta^{\mu}\theta^{-\mu},
\label{F-theta-mu}
\end{eqnarray}
we require the positivity of the inner product:
\begin{equation}
 (F,F) = \int \zeta F^* \vee F \Bigr|_\zform \vol \geq 0,
\end{equation}
or equivalently,
\begin{eqnarray}
 0 &\leq& \zeta F^* \vee F \Bigr|_\zform  \nn \\
 &=& f_0^*f_0 - (\alpha\delta + \beta\gamma)
       \left\{ T_{-\mu}(f_+^* f_+) + T_{\mu}(f_-^* f_-)\right\} \nonumber\nn\\
 && {}- (\alpha\gamma + \beta\delta)
       \left\{ T_{-\mu}(f_+^* f_-) \Tright{-2\mu}
               + T_\mu (f_-^* f_+) \Tright{2\mu} \right\} \nn \\
 && {}- (\alpha^2 - \beta^2)(\gamma^2 - \delta^2) \tilde{f}^* \tilde{f}.
 \end{eqnarray}
This relation then requires the following constraints on the constant 
parameters:
\begin{eqnarray}
 \alpha\delta + \beta\gamma \leq 0, && \qquad \alpha\gamma+\beta\delta = 0
 \nn \\
 (\alpha^2 - \beta^2)(\gamma^2 - \delta^2)
  &=& (\alpha\gamma + \beta\delta)^2 - (\alpha\delta + \beta\gamma)^2 \leq 0. 
\end{eqnarray}
We can then summarize the constraints required from the positivity of 
the inner product as
\begin{equation}
 \alpha\delta + \beta\gamma = -r , \qquad \alpha\gamma + \beta\delta = 0
\end{equation}
with $r \geq 0$.

We further impose that $dx^\mu\vee{}$ should satisfy the same Clifford
algebra as the continuum case,
\begin{eqnarray}
\{dx^\mu,dx^\nu\}_\vee &=&
 dx^\mu \vee dx^\nu + dx^\nu \vee dx^\mu \nn \\
   &=& (\theta^\mu - \theta^{-\mu}) \vee (\theta^\nu - \theta^{-\nu})
     + (\theta^\nu - \theta^{-\nu}) \vee (\theta^\mu - \theta^{-\mu})\nn \\
   &=& 4 r \delta^{\mu\nu} = 2\delta^{\mu\nu},
\end{eqnarray}
which determines the parameter $r=\frac{1}{2}$.
With these parameters the inner product is given by
\begin{eqnarray}
&&(f\theta^{\epsilon_1 \mu_1} \ldots \theta^{\epsilon_p \mu_p}, 
g \theta^{\epsilon'_1 \nu_1} \ldots \theta^{\epsilon'_q \nu_q}) \nn \\
&=&
 \lefteqn{
  \int \left\{
  \zeta f \theta^{\epsilon_1 \mu_1} \ldots \theta^{\epsilon_p \mu_p}\right\}^*
   \vee
         g \theta^{\epsilon'_1 \nu_1} \ldots \theta^{\epsilon'_q \nu_q}
   \Bigr|_\zform \vol
  }\hspace{1em}\nonumber\\
 &=& \int  \frac{1}{2^p} \delta_{pq}
     \left( T_{\epsilon_1\mu_1 \ldots \epsilon_p\mu_p}f^*g \right)
     \delta^{\epsilon_1\mu_1}_{[\epsilon'_1\nu_1} \ldots
            \delta^{\epsilon_p\mu_p}_{\epsilon'_p \nu_p]} \vol \nn \\
 &=& \sum_x  \frac{1}{2^p} \delta_{pq}
     \delta^{\epsilon_1\mu_1}_{[\epsilon'_1\nu_1} \ldots
            \delta^{\epsilon_p\mu_p}_{\epsilon'_p \nu_p]} f_x^* g_x,
\end{eqnarray}
which is the general form of (\ref{eq:inner-p-Dimakis}).

The final form of our new Clifford product is
\begin{eqnarray}
 f \theta^K &\vee& g \theta^L
  = \sum_{p=0}^{2D} (-)^{p(p-1)/2} \left(-\frac{1}{2}\right)^p
        \sum_{\{\epsilon_i \mu_i\}}\nn\\*
  && \left\{ \eta^p \left(\prod_{i=1}^p \ddx{\theta^{\epsilon_i \mu_i}}
                             {T}_{\epsilon_i \mu_i}\right)
                f \theta^K \right\}
	\wedge \left\{ \left(\prod_{i=1}^p \ddx{\theta^{-\epsilon_i \mu_i}}
	                         \right)
                       g \theta^L \right\},
\label{s-clifford-product}                       
\end{eqnarray}
where it should be understood that permutably equivalent product of the 
differentiation of $\theta^{\pm\mu}$ should not be summed up in the 
summation.
The reason why we have shift operator without arrow $T_{\epsilon_i \mu_i}$ 
in this definition is due to the cancellation of the operation of 
$\Tright{\epsilon_i\mu_i}$ and $\Tright{-\epsilon_i\mu_i}$ and 
thus the shift operator 
$T_{\epsilon_i \mu_i}$ acts only on the fields restricted in the range. 

Let us summarize this section.
We have postulated the form of the new Clifford product on the symmetric 
lattice where we have introduced 4 parameters but there are only two 
actual degrees of freedom, $\alpha\delta +
\beta\gamma$ and $\alpha\gamma + \beta\delta$. 
The associativity of Clifford product had no restriction on the parameters 
while the positivity restricted the parameters to single degree of freedom.
The remaining freedom of the parameter has been fixed by the isomorphism 
between the Clifford algebra of $\gamma$-matrix and 
1-form Clifford products.
Thus the newly defined Clifford product is almost ``unique'' on the symmetric 
lattice as far as $2D$ forms are introduced with noncommutative differential 
forms.

It is important to note that the summation $\sum_{p=0}^{2D}$ in the 
definition of the Clifford product (\ref{s-clifford-product}) runs over 
$p=0\sim 2D$. Correspondingly the summation $\sum_{\{\epsilon_i \mu_i\}}$ 
sums up $\pm\mu$ for $\mu=1\sim D$. 
This is due to the fact that we have $2D$ 1-forms 
$\theta^{\pm\mu}$ on the symmetric lattice. 
Restricting the above summation to $p=0\sim D$, we can define another 
type of Clifford product which satisfies associativity, 
\begin{eqnarray}
 f \theta^K \vee g \theta^L 
  &\equiv& \sum_{p=0}^{D} (-)^{p(p-1)/2}\sum_{\{\mu_i\}}\frac{1}{p!}
     \left\{ \eta^p \prod_{i=1}^p \left(
        \alpha \ddx{\theta^{\mu_i}} \Tright{\mu_i}
        + \beta \ddx{\theta^{-\mu_i}} \Tright{-\mu_i}
     \right)\right\} f \theta^K \nonumber\\
  &&
   \hspace{3em}
   {}\wedge \left\{ \prod_{i=1}^p \left(
        \gamma \ddx{\theta^{\mu_i}} \Tright{\mu_i}
        + \delta \ddx{\theta^{-\mu_i}} \Tright{-\mu_i}
     \right)\right\} g \theta^L,
  \label{eq:Clifford-2-def}
\end{eqnarray}
where we have again introduced four parameters $\alpha, \beta, \gamma$ 
and $\delta$. Here it should be noted that the summation runs over 
only $+\mu$ for $\mu_i$. 
The proof of the associativity of this Clifford product is given in 
appendix D.

In defining the inner product of (\ref{inner-product}), we employ the 
above definition of the Clifford product. We impose the positivity of 
a self inner product of arbitrary field $\Phi$ as in 
(\ref{self-positivity}).
Since the formula (\ref{factorization1}) characterizing the factorization 
property of Clifford product is still satisfied in the present case, 
the positivity requirement of the inner product is equivalent to the 
following condition:
\begin{eqnarray}
 0 &\leq& \zeta F^* \vee F \Bigr|_\zform  \nn \\
 &=& f_0^*f_0 - \beta\gamma T_{-\mu} (f_+^* f_+) 
     - \alpha\delta T_\mu (f_-^* f_-) 
     \nonumber\\
 &&  {}- \beta\delta T_{-\mu}(f_+^* f_-)  \Tright{-2\mu}
     - \alpha\gamma T_\mu (f_-^* f_+) \Tright{2\mu}
\label{another-clifford-product-original}     
\end{eqnarray}
where $F$ has the same form as (\ref{F-theta-mu}).
This positivity constraint limits the parameters as follows:
\begin{equation}
 \beta\gamma \leq 0, \ \ \ \alpha\delta \leq 0, \ \ \ 
  \beta\delta = \alpha\gamma = 0.
\end{equation}
Non-trivial solutions on these relations are
\begin{enumerate}
 \item $\alpha=\delta=0, \beta\gamma < 0 $,
 \item $\beta=\gamma=0,  \alpha\delta < 0$,
\end{enumerate}
where the former picks up only $f_+^* f_+$ term while the latter 
picks up $f_-^* f_-$term in the factorized part of the inner product.

Taking the parameter choice of the case 1. with $\beta\gamma=-1$ in the 
solutions, we obtain another concrete definition of Clifford product on the 
symmetric lattice  
\begin{eqnarray}
 && f \theta^K \vee g \theta^L \nn \\
  &&= \sum_{p=0}^{D} (-)^{p(p-1)/2} (-)^p 
        \sum_{ \{\mu_i\}} \frac{1}{p!}
	\left\{ \eta^p \left(\prod_{i=1}^p \ddx{\theta^{- \mu_i}}
                             T_{- \mu_i}\right)
                f \theta^K \right\}
	\wedge \left\{ \prod_{i=1}^p \ddx{\theta^{ \mu_i}}
                       g \theta^L \right\}.
                       \qquad\quad
\label{another-symmetric-clifford-product}
\end{eqnarray}
As an particular example the 1-form Clifford product effectively 
has the following asymmetric operation:
\begin{equation}
\theta^\mu \vee = \theta^\mu \wedge, \ \ \ 
\theta^{-\mu} \vee = \theta^{-\mu} \wedge - \ddx{\theta^\mu}.
\end{equation}

It is important to note that $\delta = d - d\vee{}$ plays a role of 
adjoint differential operator of $d$ as in the continuum case:
\begin{equation}
 \int (\zeta g \theta^K )^* \vee df \theta^L \Bigr|_\zform \vol
  = \int (\zeta \delta g \theta^K )^* \vee f \theta^L \Bigr|_\zform \vol,
\end{equation}
where the proofs based on the definitions of the Clifford products 
(\ref{s-clifford-product}) and (\ref{another-symmetric-clifford-product}) 
are given in appendix E. In both definitions of the Clifford product 
the operator $\delta$ plays the role of the conjugate of the exterior 
derivative operator. 
This shows that differential form and Clifford product is well defined on the 
symmetric lattice.


\section{Dirac-K\"ahler Fermion and Staggered Fermion on the Symmetric Lattice}


\setcounter{equation}{0}

One of the puzzling points in formulating differential form on the 
$D$-dimensional symmetric lattice is that we have $2^{2D}$ differential 
forms since we need $2D$ 1-forms $\theta^\mu$ and $\theta^{-\mu}$ or 
$dx^\mu$ and $\tau^\mu$ to accommodate Hermiticity. 
However we just need  $2^D$ forms in formulating Dirac-K\"ahler fermion 
formulation in general since there are $D$ 1-forms in $D$-dimensional 
manifold. 

In this section we show that a naive choice of $2^D$ forms for the 
Dirac-K\"ahler fermion 
formulation with the Clifford product defined in the previous section 
naturally leads to the Kogut-Susskind fermion 
action\cite{Kogut:1975ag,Susskind:1977jm}.
Here in this section we mainly deal with the Clifford product 
defined in (\ref{s-clifford-product}).
We can then show that the Dirac-K\"ahler fermion formulation with a 
particular choice of the form degrees for fermion and the conjugate 
fermion on the symmetric lattice leads to the staggered fermion action 
which is commonly used in lattice 
QCD\cite{Kawamoto:1981hw,Kluberg-Stern:1983dg,Gliozzi:1982ib}. 

\subsection{Naive Fermion}

We first consider the case where we choose $dx^\mu$ and $\tau^\mu$ 
as the basis of fields separately and thus define 
\begin{eqnarray}
 Z^i{}_{\!(j)}
  &=& \sum_{p=0}^D \frac{1}{p!}
        \left( \gamma_{\mu_1}^T \gamma_{\mu_2}^T
	         \ldots \gamma_{\mu_p}^T \right)^i{}_{\!(j)}\,\,
	dx^{\mu_1} \wedge dx^{\mu_2} \wedge \cdots \wedge dx^{\mu_p},
 \label{eq:Z-naive}	\\
 W^i{}_{\!(j)}
  &=& \sum_{p=0}^D \frac{1}{p!}
        \left( \gamma_{\mu_1}^T \gamma_{\mu_2}^T
	         \ldots \gamma_{\mu_p}^T \right)^i{}_{\!(j)}\,\,
	\tau^{\mu_1} \wedge \tau^{\mu_2} \wedge \cdots \wedge \tau^{\mu_p},	
\label{eq:W-naive}
\end{eqnarray}
where we rewrite the definitions of $dx^\mu$ and $\tau^\mu$:
\begin{equation}
dx^\mu = \theta^\mu - \theta^{-\mu},\ \ \ 
\tau^\mu = \theta^\mu + \theta^{-\mu}.
\end{equation}
We can then derive the following relations which are the essence of the 
Dirac-K\"ahler formulation:
\begin{eqnarray}
 dx^\mu \vee Z^i{}_{\!(j)}
  &=& (\gamma_\mu^T Z)^i{}_{\!(j)}, \ \ \ 
 \tau^\mu \vee Z^i{}_{\!(j)}
  = \tau^\mu \wedge Z^i{}_{\!(j)},\nn \\
 dx^\mu \vee W^i{}_{\!(j)}
  &=&  dx^\mu \wedge W^i{}_{\!(j)}, \ \ \ 
 \tau^\mu \vee W^i{}_{\!(j)}
  = (\gamma_5^{\dagger T} W (\gamma_5 \gamma_\mu)^T)^i{}_{\!(j)}. 
\end{eqnarray}
There are the following orthogonality relations for the bases:
\begin{eqnarray}
\left( f Z^{k}{}_{(l)}, g Z^i{}_{(j)}\right) 
&=& \left(f W^{k}{}_{(l)}, g W^i{}_{\!(j)}\right) 
= 2^{D/2}\, \delta^{(l)}_{(j)}\, \delta^i_k\, \sum_x f^*(x) g(x),\nn \\ 
\left( f Z^{k}{}_{(l)}, g W^i{}_{\!(j)}\right) 
&=& \left(f W^{k}{}_{(l)}, g Z^i{}_{\!(j)}\right) 
=  \delta^{(l)}_{k}\, \delta^i_{(j)}\, \sum_x f^*(x) g(x), 
\end{eqnarray}
where the *-operation on the bases leads:
\begin{equation}
\left(Z^{k}{}_{(l)}\right)^* = \zeta Z^{(l)}{}_{k}, \ \ \ 
\left(W^{k}{}_{(l)}\right)^* = \zeta\eta W^{(l)}{}_{k},
\end{equation}
where the sign operators $\zeta$ and $\eta$ are defined in 
(\ref{sign-factor-zeta}) and (\ref{sign-factor-eta}), respectively.
We can further show the following relations:
\begin{eqnarray}
 \left(f Z^{k}{}_{(l)}, g dx^\mu \vee W^i{}_{\!(j)}\right)
&=& (\gamma_\mu^T)^{(l)}{}_k\delta^i{}_{\!(j)}\sum_x f^*(x) g(x), \nn \\
 \left(f W^{k}{}_{(l)}, g \tau^\mu \vee Z^i{}_{\!(j)}\right)
&=& (\gamma_\mu^T)^{(l)}{}_k\delta^i{}_{\!(j)}\sum_x f^*(x) g(x),
\end{eqnarray}
where we have used the following relations:
\begin{eqnarray}
 W^{(l)}{}_{k} \vee \tau^\mu &=& 
 -((\gamma_5 \gamma_\mu)^T W (\gamma_5^\dagger)^T)^{\!(l)}{}_k \\
 Z^{\!(l)}{}_k \vee dx^\mu &=& (Z \gamma_\mu^T)^{\!(l)}{}_k.
\end{eqnarray}

We rewrite the exterior derivative operator defined in 
(\ref{eq:differential}) as
\begin{eqnarray}
 d 
  &=& \sum_\mu\left[\theta^\mu \del_{-\mu} - \theta^{-\mu} \del_{+\mu}\right]
  \nn \\
  &=& \sum_\mu\left[ dx^\mu \nabla_\mu - \tau^\mu \Delta_\mu\right],
\end{eqnarray}
where
\begin{eqnarray}
\nabla_\mu &=& \frac{1}{2}(\del_{+\mu} + \del_{-\mu}), \nn \\
\Delta_\mu &=& \frac{1}{2}(\del_{+\mu} - \del_{-\mu}).
\end{eqnarray}
Here $\nabla_\mu$ is the lattice derivative and $\Delta_\mu$ is 
the lattice second derivative defined in (\ref{def-difference-operator}). 
We then expand the fermionic field and the conjugate fermionic field 
composed of differential forms in terms of the bases $Z$ and $W$,
\begin{eqnarray}
\Psi = \psi_i{}^{(j)}Z^i{}_{\!(j)},&& \ \ \  \Psibar = 
(\zeta Z^{k}{}_{(l)})^*\psibar_{(l)}{}^k = Z^{(l)}{}_k
\psibar_{(l)}{}^k, \nn \\
\chi = \chi_i{}^{(j)}W^i{}_{\!(j)},&& \ \ \ 
\overline{\chi} = (\zeta W^{k}{}_{(l)})^*\overline{\chi}_{(l)}{}^k = 
\eta W^{(l)}{}_k\overline{\chi}_{(l)}{}^k. 
\label{def-fermion-basis-1}
\end{eqnarray}

We can now explicitly construct the Dirac-K\"ahler fermion action:
\begin{eqnarray}
\int&& \left(
\begin{array}{cc}
\Psibar \vee d \vee \Psi, & \Psibar \vee d \vee \chi \\
\overline{\chi} \vee d \vee \Psi, & \overline{\chi} \vee d \vee \chi
\end{array}
\right) \Bigr|_\zform \vol  \nonumber \\[.3em]
=&& \sum_x \left(
\begin{array}{cc}
\hspace{-1em}
2^\frac{D}{2}\sum\limits_\mu \overline{\psi}_{(j)}\gamma^\mu\nabla_\mu\psi^{(j)}, & 
\hspace{-3.5em}
\sum\limits_\mu \{ \tr(\overline{\psi}\gamma_\mu)\nabla_\mu\tr\chi-\tr\overline{\psi}
\Delta_\mu\tr(\gamma_\mu\chi) \} \\
\sum\limits_\mu \{(\tr\overline{\chi})\nabla_\mu\tr(\gamma_\mu\psi)+
\tr(\overline{\chi}\gamma_\mu)\nabla_\mu\tr\psi\},& 
2^\frac{D}{2}\sum\limits_\mu((\gamma_5\gamma_\mu\overline{\chi})_{(j)}
\Delta_\mu(\gamma_5^\dagger\chi)^{(j)})
\end{array}
\right). \nn \\
\end{eqnarray} 
In particular we obtain the following Kogut-Susskind fermion action:
\begin{eqnarray}
 \lefteqn{
 \int \left( \Psibar \vee (d + m) \vee \Psi \right)
      \Bigr|_\zform \vol
  }
  \nn\\*
 &=& 2^{D/2} \sum_x \left[
      \sum_\mu \left\{
       \psibar_{(j)}{}^i(x) (\gamma^\mu){}_i{}^k 
          \nabla_\mu \psi_k^{(j)}(x)
       \right\}
       + m \psibar_{(j)}{}^i(x) \psi_i{}^{(j)}(x) \right]
\label{eq:naive-action}
\end{eqnarray}
where the fermion fields are expanded only by the differential forms of 
$dx^\mu$. 
This naive truncation of $2^{2D}$ forms into $2^{D}$ form leads the 
naive lattice fermion action with $2^{\frac{D}{2}}$ flavors.
To be precise the above lattice Dirac action is not the original 
Kogut-Susskind fermion action since the fields on the lattice and the 
components of Dirac fields have no direct correspondence. 
We, however, consider that the correspondence is hidden through the 
$n$-form of differential form and $n$-simplex of lattice component. 

At this stage it is worth to mention that we have chosen the bases of $2^D$ 
possible differential forms as purely composed of $dx^\mu$ or $\tau^\mu$. 
There are in general mixed bases of $dx^\mu$ and $\tau^\mu$ which we have 
not considered in this naive example.

\subsection{Staggered Fermion}

We next consider the case where $\Psi$ is expanded by $\theta^{\mu_i}$ 
while $\Psibar$ is expanded by $\theta^{-\mu_i}$.
We introduce the following basis for $\Psi$:
\begin{equation}
\Psi = \psi_i{}^{(j)} Z_+{}^i{}_{\!(j)},
\label{fermion-basis-2}
\end{equation}
where
\begin{eqnarray}
 Z_+{}^i{}_{\!(j)}
  &=& \sum_{p=0}^D \frac{1}{p!}
       (\gamma_{\mu_1}^T \ldots \gamma_{\mu_p}^T)^i{}_{\!(j)}
        \sqrt{2}\theta^{\mu_1} \ldots \sqrt{2}\theta^{\mu_p}.
\end{eqnarray}
The actions of $dx^\mu \vee{}$ and $\tau^\mu \vee{}$ on this basis are 
given by 
\begin{eqnarray}
 (\theta^\mu - \theta^{-\mu}) \vee Z_+{}^i{}_{\!(j)}
  &=& \frac{1}{\sqrt{2}}(\gamma_\mu^T Z_+ )^i{}_{\!(j)}
       - \theta^{-\mu} Z_+{}^i{}_{\!(j)}, \nn \\
 (\theta^\mu + \theta^{-\mu}) \vee Z_+{}^i{}_{\!(j)}
  &=& \frac{1}{\sqrt{2}}
       (\gamma_5^\dagger{}^T Z_+ (\gamma_5 \gamma_\mu)^T)^i{}_{\!(j)}
       + \theta^{-\mu} Z_+{}^i{}_{\!(j)}.
\end{eqnarray}

Similarly we expand $\Psibar$ by the following bases: 
\begin{equation}
\Psibar = Z_-{}^{(l)}{}_k \psibar_{(l)}{}^k
\label{fermion-basis-3}
\end{equation}
where
\begin{eqnarray}
 Z_-{}^{(l)}{}_k = (\zeta Z_+{}^{k}{}_{(l)})^*
  = \sum_{p=0}^D \frac{1}{p!}
     (\gamma_{\mu_1}^T \ldots \gamma_{\mu_p}^T)^{(l)}{}_k
     (-\sqrt{2}\theta^{-\mu_1}) \ldots (-\sqrt{2}\theta^{-\mu_p}).
\end{eqnarray}
We can derive the following relations for the inner products:
\begin{eqnarray}
  \left( f Z_+{}^{k}{}_{(l)}, g Z_+{}^i{}_{\!(j)} \right)
 &=& 2^{D/2} \delta^{(l)}_{(j)} \delta^i_k  \sum_x f^*(x)g(x), \nn \\
  \left( f Z_+{}^{k}{}_{(l)}, g \theta^\mu \vee Z_+{}^i{}_{\!(j)} \right)
 &=&
 2^{D/2} \frac{1}{2\sqrt{2}}
      \left\{ \gamma_5^\dagger{}_k{}^i (\gamma_5 \gamma_\mu)_{(j)}{}^{(l)}
              + (\gamma_\mu){}_k{}^i \delta^{(l)}{}_{\!(j)} \right\}
      \sum_x f^*(x) g(x), \nn \\
 \left( f Z_+{}^{k}{}_{(l)}, g \theta^{-\mu} \vee Z_+{}^i{}_{\!(j)} \right)
 &=& 
 2^{D/2} \frac{1}{2\sqrt{2}}
       \{(\gamma_5^\dagger)_k{}^i(\gamma_5 \gamma_\mu)^{(l)}{}_{(j)} - 
       (\gamma_\mu)_k{}^i \delta^{(l)}{}_{(j)} \} 
       \sum_x f^*(x) g(x), \nn \\   
&& \left( f Z_+{}^{k}{}_{(l)}, g \theta^{-\mu} Z_+{}^i{}_{\!(j)} \right)
= 0.     
\label{eq:Zpm-orthogonal}
\end{eqnarray}

\begin{figure}
 \includegraphics{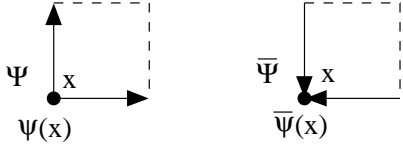}
 \caption{$\psi(x)$ on the starting point of positively oriented lattice, 
$\psibar(x)$ on the end point of negatively oriented lattice
 .}
 \label{fig:Z-pm}
\end{figure}

It is interesting to note that $\Psi$ and $\Psibar$ can be considered 
as fermion fields located on two types of oriented lattice separately; 
$\Psi$ is defined on a positively oriented lattice while $\Psibar$ 
is defined on a negatively oriented lattice. 
The Hermiticity of Dirac-K\"ahler action required the necessity of symmetric 
lattice which naturally led to the introduction of two types of 1-forms: 
$\theta^\mu$ and $\theta^{-\mu}$. 
We need to introduce the fermion field $\psi_i{}^{(j)}$ and the 
conjugate fermion field $\psibar_{(j)}{}^i$ on a lattice in the 
Dirac-K\"ahler fermion formulation. 
Correspondingly it is natural to introduce the separate conjugate 
base $Z_+$ and $Z_-$ for $\psi_i{}^{(j)}$ and $\psibar_{(j)}{}^i$, 
respectively. 
This conjugate feature can be pictorially depicted by Fig.\ref{fig:Z-pm}, 
where $\Psi(x)$ lives on the starting point of positively oriented link 
while $\Psibar(x)$ lives on the end point of negatively oriented link. 

We can now construct the action of Dirac-K\"ahler fermion by 
using the orthogonal relations (\ref{eq:Zpm-orthogonal}),
\begin{eqnarray}
 \lefteqn{
  \int  \Psibar \vee ( \sqrt{2} d + m ) \vee \Psi \Bigr|_\zform \vol
 }\hspace{1em}\\
 &=& 2^{D/2}\sum_x \biggl[
     \sum_\mu \Bigl\{
      \psibar_{(j)}{}^k(x) (\gamma_\mu){}_k{}^i
         \nabla_\mu\psi_i{}^{(j)}(x)\nn\\*[-.5em]
 && \hspace{4.5em}
     {}+ \psibar_{(j)}{}^k(x) (\gamma_5^\dagger){}_k{}^i
         \Delta_\mu\psi_i{}^{(l)}(x)
          (\gamma_5 \gamma_\mu)_{(l)}{}^{\!(j)}
     \Bigr\}\nn\\*
 && \hspace{3em}
    {}+ m \psibar_{(j)}{}^i(x)\, \psi_i{}^{(j)}(x)
   \biggr],
    \label{eq:staggered-action}
\end{eqnarray}
where the origin of the factor in $\sqrt{2} d \vee {}$ comes from 
the normalization to the Laplacian operator of this Clifford product. 
This action is exactly the same as the staggered fermion action of 
(\ref{eq:staggered-action-original}).
It is interesting to recognize that the inner product defined by the 
Clifford product of this subsection bridges the fermion field $\Psi(x)$ 
defined on the positively oriented lattice and the conjugate fermion field 
$\Psibar$ defined on the negatively oriented lattice. 

It is interesting to note at this stage that the Clifford product defined 
in (\ref{another-symmetric-clifford-product}) 
leads to the same staggered fermion formulation. 
In fact the following action leads the same staggered fermion action as 
the above (\ref{eq:staggered-action}):
\begin{equation}
  \int \left( \Psibar \vee (d + m) \vee \Psi \right)
      \Bigr|_\zform \vol,
 \label{eq:staggered-action2}    
\end{equation} 
where the Clifford product of this action is defined by 
(\ref{another-symmetric-clifford-product}) and 
the fermion fields are expanded by the naive form of (\ref{eq:Z-naive}) as
\begin{equation}
\Psi = \psi_i{}^{(j)} Z^i{}_{\!(j)}, \ \ \ 
\Psibar = Z^{(j)}\!{}_i\,\psibar_{(j)}{}^i.
\end{equation}
In this definition of the Clifford product $\sqrt{2}$ for the exterior 
derivative is not necessary. 

We claim that these are the direct derivations of staggered fermion action 
from Dirac-K\"ahler fermion action. The connection between the 
Dirac-K\"ahler fermion formulation and Kogut-Susskind fermion formulation 
was recognized at the similar early stage of lattice QCD when the staggered 
fermion formulation was recognized to be equivalent to the above expression 
of (\ref{eq:staggered-action})\cite{Kawamoto:1981hw,Kluberg-Stern:1983dg,
Gliozzi:1982ib}.
The free Dirac fermion action part of the above staggered fermion action 
may be considered as the Kogut-Susskind fermion action 
(\ref{eq:naive-action}) of previous subsection. Only difference between 
them is the choice of different bases. 

\begin{figure}[tbnh]
 \begin{center}
  \begin{minipage}{.45\linewidth}
   \includegraphics[width=\linewidth]{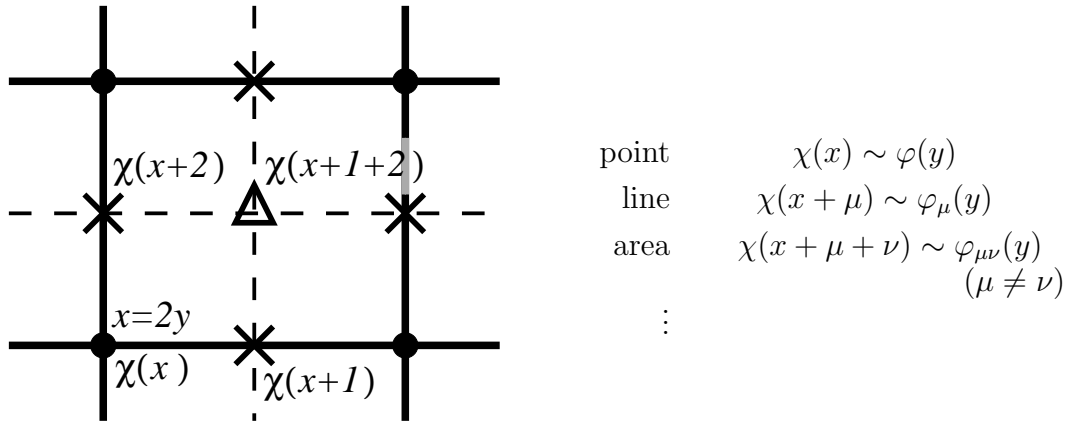}
  \end{minipage}
  \hspace{.05\linewidth}
  \begin{minipage}{.45\linewidth}
   \begin{eqnarray*}
    \mbox{point} &\displaystyle \chi(x) \sim \varphi(y) \\
    \mbox{line} &\displaystyle \chi(x+\mu) \sim \varphi_\mu(y)\\
    \mbox{area} &\displaystyle \quad\chi(x+\mu+\nu) \sim  \varphi_{\mu\nu}(y)
                \\[-.3em] & \hspace{9em} \big( \mu\neq\nu\big)\\[-.5em]
      \vdots
   \end{eqnarray*}
  \end{minipage}
 \end{center} 
\caption{$n$-form and $n$-simplex correspondence between the Dirac-K\"ahler 
fermion and the staggered fermion fields. The solid lines denote double 
spacing lattice while the dotted lines and the solid lines together compose 
of the original lattice.}
 \label{fig:staggered}
\end{figure}

The essence of the identification of Dirac-K\"ahler fermion action 
which is defined purely by a differential form and the staggered 
fermion formulation with species doublers which is equivalent to the 
naive Dirac fermion action is based on the one to one correspondence 
between $n$-form of differential algebra and $n$-simplex of simplicial 
manifold. 
For the cubic lattice as a simplicial manifold this identification 
can be realized by doubling the size of lattice. 
The field defined on a site of original lattice can be identified as a 
field defined on the center of $n$-simplex of the double spacing lattice 
and thus can be identified as a field of higher differential form. 
For example the fields on the middle of links of double spacing lattice 
correspond to 1-form fields, the fields on the middle of plaquette 
to 2-form fields, and so on. See Fig. \ref{fig:staggered}. 


\section{Yang-Mills and Dirac-K\"ahler Matter from Clifford Product}

\setcounter{equation}{0}

\label{app:Y-M} 

\begin{figure}
 \begin{minipage}{.3\linewidth}
 \includegraphics[width=\linewidth]{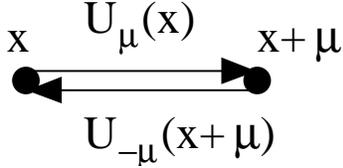}  
 \end{minipage}
 \begin{minipage}{.7\linewidth}
   \caption{$U_{-\mu}(x+\mu) = U_\mu^\dagger(x)$}
 \label{fig:u-minus}
 \end{minipage}
\end{figure}

We formulate lattice QCD with Dirac-K\"ahler matter fermion in terms 
of noncommutative differential form via Clifford product defined in the 
previous sections. Here in this section we only consider the symmetric 
lattice. Similar treatments have already been given by A.~Dimakis et al. 
\cite{Dimakis:1994qq} and P. Aschieri et al.\cite{Aschieri:2002vn}.
We define 1-form link variable $U$ 
\begin{equation}
 U = \sum_\mu \left( U_\mu \theta^\mu + U_{-\mu} \theta^{-\mu}\right),
\end{equation}
where $U_\mu$ and $U_{-\mu}$ are Lie group valued function of 
universal differential algebra defined in 
(\ref{function-of-differential-algebra}).
Hermitian conjugate of $U$ can be defined as 
\begin{eqnarray}
U^\dagger &=& \sum_\mu \left( (\theta^\mu)^* U_\mu^\dagger  + 
(\theta^{-\mu})^* U_{-\mu}^\dagger \right) \nn \\
&=& - \sum_\mu \left( T_{-\mu} U_\mu^\dagger \theta^{-\mu} + 
T_\mu U_{-\mu}^\dagger \theta^\mu\right) = - U,
\end{eqnarray}
where the link variable $U_{-\mu}$ is related to the Hermitian conjugate as 
\begin{eqnarray}
T_{-\mu} U_\mu^\dagger(x) &=& U_\mu^\dagger(x-\mu) = U_{-\mu}(x) \nn \\
T_\mu U_{-\mu}^\dagger(x) &=& U_{-\mu}^\dagger(x+\mu) = U_\mu(x),
\end{eqnarray}
which is pictorially shown in Fig. \ref{fig:u-minus}.

A gauge field is defined as
\begin{equation}
 A = \sum_\mu\left[ A_\mu \theta^\mu + A_{-\mu} \theta^{-\mu}\right] = U-u 
   = \sum_\mu\left[ (U_\mu-1) \theta^\mu + (U_{-\mu}-1) \theta^{-\mu}\right],
\label{eq:def-A}
\end{equation}
where 
\begin{equation}
u = \sum_\mu (\theta^\mu + \theta^{-\mu}).
\end{equation}
It is interesting to note that $u$ has the following special properties:
\begin{equation}
u^2 = 0, \ \ \ \ 
[u,\omega_p]_{\pm} \ = \  u \omega_p - (-)^p \omega_p u \ = \  d \omega_p,
\label{u-property}
\end{equation}
where $\omega_p$ is $p$-form variable and $[\ \ , \ \ ]_{\pm}$ is the 
graded commutator. 

A gauge transformation of the link variable $U$ is defined by
\begin{equation}
 U \longrightarrow G U G^{-1},
  \label{eq:trans-U}
\end{equation}
where $G$ is a Lie group valued $0$-form and is unitary.
In terms of components the gauge transformations of the link variable 
can be read as
\begin{eqnarray}
 U_\mu(x) &\longrightarrow& G(x)U_\mu(x) G^{-1}(x+\mu),\label{eq:trans-U-mu}\\
 U_{-\mu}(x+\mu)
   = U_\mu^\dagger(x) &\longrightarrow& G(x+\mu) U_\mu^\dagger(x) G^{-1}(x),
  \label{eq:trans-Ud-mu}
\end{eqnarray}
which are the standard lattice gauge transformation of link variables. 

From (\ref{eq:def-A}) and (\ref{u-property}) we obtain the
transformation of $A$
\begin{equation}
 A \longrightarrow GAG^{-1} + G\,dG^{-1}.
\label{eq:trans-A}
\end{equation} 
For an infinitesimal gauge transformation $G = 1 - v$ the gauge transformation 
leads to the following standard form:
\begin{equation}
   \delta_g A = dv + [A,v],
\label{1-form-gauge-transformation}   
\end{equation}
whose component-wise expression is
\begin{equation}
  \delta_g A_\mu(x)
    = \del_{+\mu} v(x) + A_\mu(x) v(x+\mu) - v(x) A_\mu(x).
\end{equation}
It should be noted that the gauge transformation of the coordinate dependent 
gauge field cannot be written down by the form of commutator due to the 
noncommutativity while the 1-form gauge field function of 
universal differential algebra has the standard transformation property.
It is worth to mention at this stage that the link variable, 
$U_\mu(x)$, and the gauge field of the standard lattice gauge theory, 
$A_\mu'(x)$, and the current definition of the gauge field $A_\mu(x)$ are 
related as $U_\mu(x)=\exp{aA_\mu'(x)}=1+A_\mu(x)$, where we have recovered the 
lattice constant $a$ while we take $a=1$ in the current definition.  
Thus the current definition of the gauge field, $A_\mu(x)$, and the 
lattice gauge field $A_\mu'(x)$ are different in the sense that 
$A_\mu(x)$ includes all the higher order terms with respect to the lattice 
constant and is not Lie algebra valued variable.  

The curvature $2$-form $F$ is equivalent to 
\begin{equation}
 U^2 = (u+A)^2 = [u,A]_+ + A^2 =  dA + A^2 \equiv F,
\end{equation}
where the special properties of $u$ in (\ref{u-property}) are used. 
The lattice version of Yang-Mills action is then given by 
\begin{eqnarray}
 \frac{1}{2g^2}\tr (F,F)
  &=& \frac{1}{2g^2}\int \tr [(\zeta F)^\dagger \vee F] 
  \Bigr|_\zform \vol \nonumber\\
  &=& \frac{1}{2g^2} \sum_x \sum_{\mu,\nu} \tr \Bigl\{
          1 - U_\mu(x) U_\nu(x+\mu) U_\mu^\dagger(x+\nu) U_\nu^\dagger(x)
          \Bigr\},\hspace{2em}
 \label{eq:Yang-Mills}
\end{eqnarray}
where we have used the definition of the Clifford product of 
(\ref{s-clifford-product}). It is interesting to note that 
the other definition of the Clifford product 
(\ref{another-symmetric-clifford-product}) leads to the same action. 
This is exactly the Wilson action of the lattice QCD. 

We now generate the gauge and Dirac-K\"ahler fermion coupled system by 
introducing a covariant derivative.
From the transformation (\ref{eq:trans-A}) and the Leibniz rule, 
we can realize that the covariant derivative has the standard form $d+A$. 
The explicit form of the covariant derivative acting on a fermion field is 
given by 
\begin{eqnarray}
 (d+A)\psi_i{}^{(j)}(x)
 &=& \psi_i{}^{(j)}(x) u + U \psi_i{}^{(j)}(x) \nn \\
 &=& \lefteqn{ -\sum_\mu
      \Bigl[ \bigl\{ U_\mu(x) \psi_i{}^{(j)}(x+\mu) - 
      \psi_i{}^{(j)}(x)\bigr\} \theta^\mu
     }\nonumber\\[-.5em]
 && \hspace{4em}
         {} + \bigl\{ U_\mu^\dagger(x-\mu) \psi_i{}^{(j)}(x-\mu) -
         \psi_i{}^{(j)}(x) \bigr\} \theta^{-\mu} \Bigr],
\end{eqnarray}
which shows obviously the covariance of the covariant derivative 
because of the transformation properties of the link variables 
(\ref{eq:trans-U-mu}) and (\ref{eq:trans-Ud-mu}) and the following 
gauge transformation of fermion field:
\begin{equation} 
\Psi \rightarrow G\Psi, \ \ \ \  \Psibar \rightarrow \Psibar G^{-1}.
\end{equation}

We can then introduce the gauge interaction to Dirac-K\"ahler matter fermion 
of the naive fermion action of (\ref{eq:naive-action}) and the staggered 
fermion action (\ref{eq:staggered-action}) and (\ref{eq:staggered-action2}).
Replacing $d$ to $d+A$ in (\ref{eq:naive-action}), we obtain 
the following gauge invariant action:
\begin{eqnarray}
 \lefteqn{
 \int \left( \Psibar \vee (d + A + m) \vee \Psi \right)
      \Bigr|_\zform \vol
  }\hspace{1em}\nn\\*
 &=& 2^{D/2} \sum_x \Biggl[
      \sum_\mu \left\{
       \psibar_{(j)}{}^i(x) (\gamma^\mu){}_i{}^k 
          \frac{1}{2}\bigl(U_\mu(x)\psi(x+\mu) 
                   - U_\mu^\dagger(x-\mu)\psi(x-\mu) \bigr)_i{}^{(j)}
       \right\}\nonumber\\[-.3em]
  && \hspace{5em}
       {}+ m \psibar_{(j)}{}^i(x) \psi_i{}^{(j)}(x) \Biggr],
\end{eqnarray}
where $\Psi$ and $\Psibar$ are defined in (\ref{def-fermion-basis-1}).
This is the naive fermion version of lattice QCD action with $2^{D/2}$ 
flavors.  

In the similar way we obtain the following gauge invariant staggered fermion 
action from $(\ref{eq:staggered-action})$:
\begin{eqnarray}
 \lefteqn{
  \int  \Psibar \vee ( \sqrt{2}(d+A) + m ) \vee \Psi \Bigr|_\zform \vol
 }\hspace{1em} \nn \\
 &=& 2^{D/2}\sum_x \biggl[
     \frac{1}{2}\sum_\mu \Bigl\{
     \psibar_{(j)}{}^k(x+\mu) (\gamma_\mu){}_k{}^i U_\mu^\dagger(x)
           \psi(x)_i{}^{(j)}
     -\psibar_{(j)}{}^k(x) (\gamma_\mu){}_k{}^i U_\mu(x) \psi(x+\mu)_i{}^{(j)}
     \nn\\*[-.5em]
 && \hspace{4.5em}
     {}+ \psibar_{(j)}{}^k(x) (\gamma_5)^\dagger{}_k{}^i
         \bigl( U_\mu(x)\psi(x+\mu) + U_\mu^\dagger(x-\mu)\psi(x-\mu) \nn \\
 &&   \hspace{3em}            
                -2\psi(x) \bigr)_i{}^{(l)}
          (\gamma_5 \gamma_\mu)_{(l)}{}^{\!(j)}
     \Bigr\}  {}+ m \psibar_{(j)}{}^i(x)\, \psi_i{}^{(j)}(x)
   \biggr],
\end{eqnarray}
where $\Psi$ and $\Psibar$ are defined in (\ref{fermion-basis-2}) 
and (\ref{fermion-basis-3}).
It is interesting to note that the other definition of Clifford 
product (\ref{another-symmetric-clifford-product}) with the 
Dirac-K\"ahler action (\ref{eq:staggered-action2}) leads the 
same staggered fermion action as the above. There is only a minor 
normalization difference between these formulations.

\section{Conclusion and Discussions}

\setcounter{equation}{0}

We have formulated the Clifford product with noncommutative differential 
forms on the lattice which satisfies associativity. 
We have directly derived Kogut-Susskind fermion action and the staggered 
fermion action from the Dirac-K\"ahler fermion formulation defined by the 
the Clifford product. 
In order to keep the Hermiticity of Dirac-K\"ahler fermion action we need 
to choose not the oriented lattice but the symmetric lattice which in general 
has $2D$ differential 1-forms in $D$-dimensional symmetric lattice. 
In defining the Clifford product there is an arbitrariness which is related 
with the doubled number of 1-forms and can be parameterized by 4 constant 
parameters. It turned out that these parameters can be determined almost 
uniquely by the positivity of the self inner product of a field.   
There are, however, two types of definition of Clifford product depending 
on the choice of $2D$ or $D$ differential 1-forms.  
It was shown that the newly defined Clifford product with noncommutative 
differential forms generates lattice QCD gauge action together with 
Dirac-K\"ahler matter fermion. 

One of the interesting point of the current analyses is that the 
Hermiticity played an important role in choosing the compatible 
lattice structure as the symmetric lattice. 
The symmetric lattice, however, has two types of 1-forms $\theta^\mu$ 
and $\theta^{-\mu}$ or $dx^\mu$ and $\tau^\mu$ on a link of lattice, and 
thus requires $2D$ 1-forms in $D$ dimension. In the standard differential 
algebra we have $D$ differential 1-forms and thus this mismatch in number 
suggests us an extra degree of freedom hidden in this formulation. 
In the naive fermion formulation of subsection 6.1 led Kogut-Susskind 
type fermion action in the sector of $dx^\mu$ base. 
On the other hand the action has the second derivative term in 
the sector of $\tau^\mu$ base.
The degree $\theta^{-\mu}$ in contrast with $\theta^\mu$ 
was used to define the conjugate of 
fermion field in the formulation of Dirac-K\"ahler fermion field 
of subsection 6.2, which led to the staggered fermion formulation with 
first and second derivatives in the action.  
It is interesting to recognize that this first derivative and the 
second derivative nature together with the doubled degrees of 
freedom of 1-form suggests both boson and fermion degrees of freedom 
hidden in this formulation. 
We believe that the extended supersymmetry on the lattice will be 
accommodated along the line of current formulation.

There is another interesting indication that the lattice chiral 
fermion problem with the recent development of Ginsparg-Wilson relation 
\cite{Ginsparg:1982bj} could be fundamentally related with the current 
formulation of noncommutative differential form. 
The noncommutative differential form was used as a convenient tool 
in the derivation of lattice chiral anomaly from the arguments of 
index theorem\cite{Fujiwara:1999fi,Fujiwara:1999fj}.

One of the importance of formulating matter fermion field in terms of 
differential form is related with the possibility that the extension of 
the formulation into gravity on a simplicial lattice may be 
straightforward since 
the gauge and matter fermion field, and Clifford product 
are all formulated by the differential form.

%

\vskip 1cm 
\noindent{\Large \textbf{Acknowledgments}}\\
This work is supported in part by Japanese 
Ministry of Education, Science, Sports and Culture under the grant number 
13640250 and 13135201.

\vskip 1cm

\begin{center}
{\Large \textbf{Appendix}\\[0pt]}
\end{center}

\noindent
{\Large \textbf{A. Proof of Stokes Theorem on the Symmetric 
Lattice}\\[0pt]}

\renewcommand{\theequation}{A.\arabic{equation}}
\setcounter{equation}{0}

We show how the Stokes theorem can be derived by the integral defined 
on the symmetric lattice in subsection 2.2.

Boundary of $r+1$-cube is
\begin{eqnarray}
 \del C_{r+1}
   = \sum_s (-)^s \left( C_r\big|_{x^s} - C_r\big|_{x^s + \mu^s}\right),
\end{eqnarray}
where $C_r\big|_{x^s}$ is a $r$-dimensional cube dimensionally reduced from
$C_{r+1}$ with $x'^s$ fixed to $x^s$. 
In order to prove the Stokes theorem, 
we must evaluate integral of exact $r+1$ form. 
Introducing the following $r$-from:
\begin{equation}
 \omega = f \theta^{\epsilon_1\mu_1}\ldots\theta^{\epsilon_r\mu_r},
\end{equation}
we can express the exact form as
\begin{eqnarray}
 d\omega
   &=& \sum_\mu 
      \left[ \del_{+\mu}f \theta^\mu
            \theta^{\epsilon_1\mu_1}\ldots\theta^{\epsilon_r\mu_r}
           - \del_{-\mu}f \theta^{-\mu}
            \theta^{\epsilon_1\mu_1}\ldots\theta^{\epsilon_r\mu_r}
       \right].
\end{eqnarray}
Since we impose $\int \theta^\mu \theta^{-\mu} = 0$, all $\mu_i$'s are 
different and thus the integrals vanishes and Stokes' theorem is trivial 
if $\mu_i = \mu_j$.
Integration of $d\omega$ in $(r+1)$-cube leads
\begin{eqnarray}
 \int_{C_{r+1}} d\omega
  &=& (-)^r \int_{C_{r+1}}\left[
      \del_{+\mu_{r+1}} f \theta^{\epsilon_1\mu_1}
       \ldots\theta^{\epsilon_r\mu_r} \theta^{\mu_{r+1}}
      - \del_{-\mu_{r+1}} f \theta^{\epsilon_1\mu_1}
       \ldots\theta^{\epsilon_r\mu_r} \theta^{-\mu_{r+1}} \right]\nn\\
  &=& (-)^r \Biggl[
      \alpha \int_{C_r\big|_{x^{\mu_{r+1}}}}
      \bigl( f(x'+\mu_{r+1}) - f(x') \bigr)
      \theta^{\epsilon_1\mu_1}\ldots\theta^{\epsilon_r\mu_r}\nn\\*
  &&  \hspace{2em}
      {}+ (1-\alpha)\int_{C_r\big|_{x^{\mu_{r+1}} + \mu_{r+1}}}
      \bigl( f(x') - f(x'-\mu_{r+1})\bigr)
      \theta^{\epsilon_1\mu_1}\ldots\theta^{\epsilon_r\mu_r}
      \Biggr]\nn\\
  &=& (-)^r \Biggl[
      \alpha \int_{C_r\big|_{x^{\mu_{r+1}}}}
      \bigl( f(x'+\mu_{r+1}) - f(x') \bigr)
      \theta^{\epsilon_1\mu_1}\ldots\theta^{\epsilon_r\mu_r}\nn\\*
  &&  \hspace{2em}
      {}+ (1-\alpha)\int_{C_r\big|_{x^{\mu_{r+1}}}}
      \bigl( f(x'+ \mu_{r+1}) - f(x')\bigr)
      \theta^{\epsilon_1\mu_1}\ldots\theta^{\epsilon_r\mu_r}
      \Biggr]\nn\\
  &=& (-)^r \int_{C_r\big|_{x^{\mu_{r+1}}}} 
            \bigl( f(x'+\mu_{r+1}) - f(x') \bigr)
      \theta^{\epsilon_1\mu_1}\ldots\theta^{\epsilon_r\mu_r},
\end{eqnarray}
where we have integrated only  $\mu_{r+1}$'s direction of $d\omega$.
Integration of the boundary is
\begin{eqnarray}
 \int_{\del C_{r+1}} \omega
  &=& (-)^{r+1} \left( \int_{C_r\big|_{x^{\mu_{r+1}}}}
       - \int_{C_r\big|_{x^{\mu_{r+1}} + \mu_{r+1}}}\right)
       f \theta^{\epsilon_1\mu_1}\ldots \theta^{\epsilon_r\mu_r}\nn\\
  &=& (-)^{r+1} \int_{C_r\big|_{x^{\mu_{r+1}}}}
       \bigl( f(x') - f(x' + \mu_{r+1})\bigr)
       \theta^{\epsilon_1\mu_1}\ldots \theta^{\epsilon_r\mu_r}\nn\\
  &=& (-)^r \int_{C_r\big|_{x^{\mu_{r+1}}}}
       \bigl( f(x' + \mu_{r+1}) - f(x')\bigr)
       \theta^{\epsilon_1\mu_1}\ldots \theta^{\epsilon_r\mu_r}\nn\\
  &=& \int_{C_{r+1}} d\omega.
\end{eqnarray}

Thus we obtain Stokes theorem on the symmetric lattice
\begin{equation}
 \int_{C} d\omega = \int_{\del C} \omega.
\end{equation}
\\

\noindent
{\Large \textbf{B. Proof of Associativity of Clifford product on the 
Oriented Lattice}\\[0pt]}
\label{app:o-associativity}

\renewcommand{\theequation}{B.\arabic{equation}}
\setcounter{equation}{0}

In this appendix we prove the associativity of the following Clifford product
on the oriented lattice: 
\begin{equation}
 (dx^K \vee dx^M) \vee dx^L
  = dx^K \vee ( dx^M \vee dx^L),
  \label{eq:o-associativity}
\end{equation}
which guarantees the associativity of Clifford product for the general form. 
We should remind that $dx^\mu=\theta^\mu$ on the oriented lattice. 

Here we first assume that the associativity is satisfied in the case where 
$dx^M$ is 1-form $dx^\mu$
\begin{equation}
 (dx^K \vee dx^\mu) \vee dx^L
  = dx^K \vee ( dx^\mu \vee dx^L),
  \label{eq:o-associ-1}
\end{equation}
where $dx^K$ and $dx^L$ are arbitrary $k,l$-form.
Assuming the associativity of (\ref{eq:o-associ-1}) we can readily show that 
the associativity is satisfied in the case 
$dx^M=dx^\mu \vee dx^\nu$ as follows: 
\begin{eqnarray}
 \left\{ dx^K \vee (dx^\mu \vee dx^\nu) \right\} \vee dx^M
  &=& \left\{ (dx^K \vee dx^\mu) \vee dx^\nu \right\} \vee dx^M
      \nonumber\\
  &=& (dx^K \vee dx^\mu )
      \vee \left\{ dx^\nu \vee dx^M \right\} \nonumber\\
  &=& dx^K
      \vee \left\{ dx^\mu \vee ( dx^\nu \vee dx^M) \right\} \nonumber\\
   &=& dx^K
      \vee \left\{ (dx^\mu \vee dx^\nu) \vee dx^M \right\}.
\end{eqnarray}
Here the Clifford product of 1-forms leads 
\begin{equation}
dx^\mu \vee dx^\nu = dx^\mu \wedge dx^\nu + 
\Tleft{-\mu}\delta_{\mu,\nu}\Tright{\nu},
\end{equation}
and thus the associativity for $dx^M$=2-form case of 
(\ref{eq:o-associativity}) is proved. 
Proceeding in the similar way by induction and noting the following general 
relation:
\begin{equation}
dx^M \vee dx^\mu = dx^M \wedge dx^\mu + 
\Tleft{-\mu}\hbox{($(m-1)$-form)}\Tright{\mu},
\end{equation}
we can complete the proof for the general form 
of the associativity (\ref{eq:o-associativity}).

We next prove that the assumed associativity (\ref{eq:o-associ-1}) is in fact 
satisfied. We can factor out $dx^\mu$ from $dx^K$ 
\begin{equation}
 dx^K = (dx^\mu)^a dx^{K'},
\end{equation}
where $dx^{K'}$ is $k'$-form and does not contain $dx^\mu$. 
Here $a = 0,1$ and thus $(dx^\mu)^a$ can be expanded by 
\begin{equation}
(dx^\mu)^a = \delta_{a,0} + \delta_{a,1} dx^\mu. 
\label{theta-mu-a-expansion}
\end{equation}
Similarly we set $dx^L = (dx^\mu)^b dx^{L'}$, where $dx^{L'}$ 
is $l'$-form and does not contain $dx^\mu$.
The following useful formula:
\begin{eqnarray}
 (dx^\mu)^a dx^{K'} \vee (dx^\mu)^b dx^{L'}
  &=& \left\{
       \delta_{a,1} \delta_{b,1}(-)^{k'} \Tleft{-\mu}\Tright{\mu}
       + (-)^{k'b} (dx^\mu)^a (dx^\mu)^b
      \right\} (dx^{K'} \vee dx^{L'})
      \nonumber\\
  &=& (-)^{k'b} \left\{ (dx^\mu)^a \vee (dx^\mu)^b \right\}
        \left\{ dx^{K'} \vee dx^{L'} \right\},
\label{eq:app-factorization1}        
\end{eqnarray}
straightforwardly leads the desired relation,
\begin{eqnarray}
 \lefteqn{
 \left\{ (dx^\mu)^a dx^{K'} \vee dx^\mu \right\}
    \vee (dx^\mu)^bdx^{L'}
 }\hspace{.4em}\nonumber\\
  &=& (dx^\mu)^a dx^{K'} \vee 
      \left\{ dx^\mu \vee (dx^\mu)^bdx^{L'} \right\}
      \nonumber\\
  &=& \left\{ \delta_{a,0} \delta_{b,0} (-)^k dx^\mu
       + \delta_{a,0} \delta_{b,1} \Tleft{-\mu} \Tright{\mu}
       + \delta_{a,1} \delta_{b,0} (-)^k \Tleft{-\mu} \Tright{\mu} \right. 
      \nn \\ 
 &&  \left.{}+ \delta_{a,1} \delta_{b,1} \Tleft{-\mu} \Tright{\mu} dx^\mu 
 \right\}  (dx^{K'} \vee dx^{L'}). 
\end{eqnarray}
This completes the proof of the associativity (\ref{eq:o-associ-1}) and 
thus the general form of (\ref{eq:o-associativity}).
\\

\noindent
{\Large \textbf{C. Proof of Associativity of Clifford Product on the Symmetric 
Lattice}\\[0pt]}

\renewcommand{\theequation}{C.\arabic{equation}}
\setcounter{equation}{0}

\label{app:associativity}
The proof of the associativity of the Clifford product on the symmetric 
lattice proceeds parallel to the oriented lattice version of the previous 
appendix except that the calculation is more involved. 

Let us again assume that the following associativity is satisfied:
\begin{equation}
 (\theta^K \vee \theta^{\epsilon\mu}) \vee \theta^L
  = \theta^K \vee (\theta^{\epsilon\mu} \vee \theta^L),
\label{eq:app-1-associativity}
\end{equation}
where $\theta^K$ and $\theta^L$ are any $k$- and $l$-form, respectively, 
while $\theta^{\epsilon\mu}$ is $1$-form.
For any $1$-form $\theta^{\epsilon'\nu}$ we can readily prove by 
(\ref{eq:app-1-associativity}) 
\begin{eqnarray}
 \{ \theta^K \vee (\theta^{\epsilon\mu} \vee \theta^{\epsilon'\nu}) \}
      \vee \theta^L
  = \theta^K \vee
       \{ (\theta^{\epsilon\mu} \vee \theta^{\epsilon'\nu}) \vee \theta^L\}.
 \label{eq:app-1a-associativity}
\end{eqnarray}
Here the Clifford product of 1-forms leads 
\begin{eqnarray}
 \theta^{\epsilon\mu} \vee \theta^{\epsilon'\nu}
  = \theta^{\epsilon\mu} \wedge \theta^{-\epsilon'\nu}
    + (\alpha\gamma + \beta\delta) \delta_{\epsilon\mu, \epsilon'\nu}
      \Tright{2\epsilon\mu}
    + (\alpha\delta + \beta\gamma) \delta_{\epsilon\mu, -\epsilon'\nu}
\end{eqnarray}
and thus the associativity for $\theta^M$=2-form case of 
(\ref{symmetric-lattice-associativity0 }) is proved.

Proceeding in the similar way by induction and noting the following general 
relation:
\begin{equation}
\theta^M \vee \theta^{\epsilon\mu} = \theta^M \wedge \theta^{\epsilon\mu} + 
\hbox{($(m-1)$-form)},
\end{equation}
we can complete the proof for the following general form 
of the associativity as in the oriented lattice case,
\begin{equation}
(\theta^K \vee \theta^M) \vee \theta^L
  = \theta^K \vee (\theta^M \vee \theta^L),
\label{eq:app-associativity}
\end{equation}
where $\theta^M$ is general $m$-form.

We next prove the assumed associativity(\ref{eq:app-1-associativity}).
We then factor out $\theta^{\pm\mu}$ from $\theta^K$ and $\theta^L$ 
\begin{eqnarray}
 \theta^K &=& (\theta^\mu)^a (\theta^{-\mu})^b \theta^{K'}, \nn \\
 \theta^L &=& (\theta^\mu)^c (\theta^{-\mu})^d \theta^{L'}
\end{eqnarray}
where $a,b,c,d=0$ or $1$ and $\theta^{K'}$ and $\theta^{L'}$ are $k'$- and 
$l'$-form, respectively, and do not contain $\theta^{\pm\mu}$.  
The following factorization formula is useful for the proof:
\begin{eqnarray}
 \lefteqn{
 \left\{ (\theta^{\epsilon\mu})^a (\theta^{-\epsilon\mu})^b\theta^{K'}\right\}
  \vee
  \left\{(\theta^{\epsilon\mu})^c (\theta^{-\epsilon\mu})^d\theta^{L'} \right\}
 } \nonumber\\
 &=& (-)^{k'(c+d)}\Bigl[
      (\theta^{\epsilon\mu})^a (\theta^{-\epsilon\mu})^b
        (\theta^{\epsilon\mu})^c (\theta^{-\epsilon\mu})^d \nonumber\\
 && \hspace{1em}
    {}+(\alpha\delta + \beta\gamma)\left\{
      \delta_{a,1}\,\delta_{d,1}(-)^{b+c}
        (\theta^{-\epsilon\mu})^b (\theta^{\epsilon\mu})^c
     + \delta_{b,1}\,\delta_{c,1}
        (\theta^{\epsilon\mu})^a (\theta^{-\epsilon\mu})^d
    \right\}\nonumber\\
 && \hspace{1em}
    {}+ (\alpha\gamma + \beta\delta)\left\{
      \delta_{a,1}\,\delta_{c,1}(-)^b
        (\theta^{-\epsilon\mu})^b (\theta^{-\epsilon\mu})^d
         \Tright{2\epsilon\mu}
      + \delta_{b,1}\,\delta_{d,1}(-)^c
         (\theta^{\epsilon\mu})^a (\theta^{\epsilon\mu})^c
         \Tright{-2\epsilon\mu}
    \right\}\nonumber\\
 && \hspace{1em}
     {} - (\alpha^2 - \beta^2)(\gamma^2 - \delta^2)
       \delta_{a,1}\delta_{b,1}\delta_{c,1}\delta_{d,1}
\Bigr] \left( \theta^{K'} \vee \theta^{L'} \right) \nn \\
 &=& (-)^{k'(c+d)}
     \left[ (\theta^{\epsilon\mu})^a (\theta^{-\epsilon\mu})^b \vee
        (\theta^{\epsilon\mu})^c (\theta^{-\epsilon\mu})^d \right]
     \left( \theta^{K'} \vee \theta^{L'} \right),
 \label{eq:app-formula2}
\end{eqnarray}
where we have used the following relation:
\begin{equation}
(\theta^\mu)^a = \delta_{a,0} + \delta_{a,1} \theta^\mu. 
\end{equation}
Special case of this formula leads
\begin{eqnarray}
 \lefteqn{
  (\theta^{\epsilon\mu})^a (\theta^{-\epsilon\mu})^b \theta^{K'} 
     \vee \theta^{\epsilon\mu}
  }\hspace{1em}\nonumber\\
 &=& \left[
    \left\{ (\theta^{\epsilon\mu})^a (\theta^{-\epsilon\mu})^b\theta^{K'} 
    \right\}
    \vee
    \left \{(\theta^{\epsilon\mu})^c (\theta^{-\epsilon\mu})^d\theta^{L'} \right\}
    \right]_{c=1,d=0,\theta^{L'}=1} \nonumber\\
 &=& (-)^{k'} \Bigl[
      \delta_{a,0} (-)^b \theta^{\epsilon\mu}(\theta^{-\epsilon\mu})^b
       \nonumber\\
 && \hspace{2.5em}
     {}+ (\alpha\delta + \beta\gamma) \delta_{b,1} (\theta^{\epsilon\mu})^a
       \nonumber\\
 && \hspace{2.5em}
     {}+ (\alpha\gamma + \beta\delta) \delta_{a,1} (-)^b
       (\theta^{-\epsilon\mu})^b \Tright{2\epsilon\mu}
     \Bigr] \theta^{K'}
\end{eqnarray}
and
\begin{eqnarray}
 \lefteqn{
 \theta^{\epsilon\mu} \vee
   \left\{ (\theta^{\epsilon\mu})^c (\theta^{-\epsilon\mu})^d \theta^{L'} 
   \right\}
 }\hspace{.5em}\nonumber\\
 &=& \Bigl[ \delta_{c,0}\theta^{\epsilon\mu} (\theta^{-\epsilon\mu})^d
     \nonumber\\
 && \hspace{2em}
    {}+ (\alpha\delta + \beta\gamma) \delta_{d,1}(-)^c (\theta^{\epsilon\mu})^c
     \nonumber\\
 && \hspace{2em}
    {}+ (\alpha\gamma + \beta\delta) \delta_{c,1}
     (\theta^{-\epsilon\mu})^d \Tright{{2\epsilon\mu}}
    \Bigr] \theta^{L'}.
\end{eqnarray}
After straightforward calculations we finally obtain
\begin{eqnarray}
 \lefteqn{
  \left\{(\theta^{\epsilon\mu})^a (\theta^{-\epsilon\mu})^b \theta^{K'} 
    \vee \theta^{\epsilon\mu} \right\}
   \vee (\theta^{\epsilon\mu})^c (\theta^{-\epsilon\mu})^d \theta^{L'}
   }\nonumber\\
 &=& (\theta^{\epsilon\mu})^a (\theta^{-\epsilon\mu})^b \theta^{K'} 
    \vee \left\{ \theta^{\epsilon\mu}
   \vee (\theta^{\epsilon\mu})^c (\theta^{-\epsilon\mu})^d \theta^{L'} \right\}
   \nonumber\\
 &=&
  (-)^{k(c+d+1)} \Bigl[
    \delta_{a,0}\,\delta_{b,0}\,\delta_{c,0}\,\delta_{d,0} \theta^{\epsilon\mu}
    + \delta_{a,0}\,\delta_{c,0}
      ( \delta_{b,0}\,\delta_{d,1} - \delta_{b,1}\,\delta_{d,0})
      \theta^{\epsilon\mu}\theta^{-\epsilon\mu}
     \nonumber\\
 && 
    {}+(\alpha\delta + \beta\gamma)\Bigl\{
     \delta_{a,0}\,\delta_{c,0}
       ( \delta_{b,1}\,\delta_{d,0} + \delta_{b,0}\,\delta_{d,1} )
    \nonumber\\*
 && \hspace{6em}
    {}+ (\delta_{a,1}\,\delta_{b,1}\,\delta_{c,0}\,\delta_{d,0}
         - \delta_{a,0}\,\delta_{b,0}\,\delta_{c,1}\,\delta_{d,1}\,)
         \theta^{\epsilon\mu}
    \nonumber\\*
 && \hspace{6em}
    {}+ 2 \delta_{a,0}\,\delta_{b,1}\,\delta_{c,0}\,\delta_{d,1}
        \theta^{-\epsilon\mu}
    \nonumber\\*
 && \hspace{6em}
    {}+ \delta_{b,1}\,\delta_{d,1}
        ( \delta_{a,0}\,\delta_{c,1} + \delta_{a,1}\,\delta_{c,0} )
        \theta^{\epsilon\mu}\theta^{-\epsilon\mu}
    \Bigr\}
    \nonumber\\
 && 
    {}+(\alpha\gamma + \beta\delta)\Bigl\{
    \delta_{b,0}\,\delta_{d,0}
     (\delta_{a,1}\,\delta_{c,0} + \delta_{a,0}\,\delta_{c,1}\,)
     \Tright{2\epsilon\mu}
    \nonumber\\*
 && \hspace{6em}
    {}+ (\delta_{a,1}\,\delta_{b,0}\,\delta_{c,1}\,\delta_{d,0}
         \Tright{2\epsilon\mu}
       - \delta_{a,0}\,\delta_{b,1}\,\delta_{c,0}\,\delta_{d,1}
         \Tright{-2\epsilon\mu}) 
        \theta^{\epsilon\mu}
    \nonumber\\*
 && \hspace{6em}
    {}+ \bigl\{\delta_{a,0}\,\delta_{c,1}
        ( \delta_{b,0}\,\delta_{d,1} + \delta_{b,1}\,\delta_{d,0})
        + \delta_{a,1}\,\delta_{c,0}
        ( \delta_{b,0}\,\delta_{d,1} - \delta_{b,1}\,\delta_{d,0})
        \bigr\} \Tright{2\epsilon\mu} \theta^{-\epsilon\mu}
    \nonumber\\*
 && \hspace{6em}
    {}+ \delta_{a,1}\,\delta_{c,1}
        ( \delta_{b,1}\,\delta_{d,0}\, + \delta_{b,0}\,\delta_{d,1})
        \Tright{2\epsilon\mu}\theta^{\epsilon\mu}\theta^{-\epsilon\mu}
    \Bigr\} \nonumber\\
 && \hspace{6em}
    {}+ (\alpha^2 - \beta^2)(\gamma^2 - \delta^2 )\Bigl\{
      \delta_{b,1}\,\delta_{d,1}
       ( \delta_{a,0}\,\delta_{c,1} - \delta_{a,1}\,\delta_{c,0}\,)
    \nonumber\\*
 && \hspace{6em}
     {} + \delta_{a,1}\,\delta_{b,1}\,\delta_{c,1}\,\delta_{d,1}
          \theta^{\epsilon\mu}
    \Bigr\} 
  \Bigr] \left( \theta^{K'} \vee \theta^{L'} \right).
\end{eqnarray}
This completes the proof of the associativity (\ref{eq:app-1-associativity}) 
and thus the general form of (\ref{eq:app-associativity}).
\\

\noindent
{\Large \textbf{D. Other Possible Associative Clifford Products on the 
Lattice}\\[0pt]}

\renewcommand{\theequation}{D.\arabic{equation}}
\setcounter{equation}{0}
\label{app:other-Cliffords}

We prove the associativity of another type of Clifford product 
defined in (\ref{eq:Clifford-2-def}).
We follow the same procedure as the previous appendices so that 
we just need to prove the following type of associativity:
\begin{equation}
 (\theta^K \vee \theta^{\epsilon\mu}) \vee \theta^L
  = \theta^K \vee (\theta^{\epsilon\mu} \vee \theta^L).
\label{eq:app-D-associativity-1form}  
\end{equation}
Since the summation in the definition of the Clifford product 
in (\ref{eq:Clifford-2-def}) does not contain the summation of $\epsilon_i$, 
we need to show the proof for both cases of $\theta^\mu$ and $\theta^{-\mu}$ 
explicitly. Straightforward calculations lead 
\begin{eqnarray}
  \lefteqn{
 \left\{ (\theta^{\mu})^a (\theta^{-\mu})^b\theta^{K'}
  \vee
  \theta^\mu \right\} \vee
    (\theta^{\mu})^c (\theta^{-\mu})^d\theta^{L'}
 }
  \nonumber\\
 &=& (\theta^{\mu})^a (\theta^{-\mu})^b\theta^{K'}
  \vee
  \left\{ \theta^\mu \vee
    (\theta^{\mu})^c (\theta^{-\mu})^d\theta^{L'} \right\}
  \nonumber\\
 &=& (-)^{k'(c+d+1)} \Bigl[
     \delta_{a,0}\,\delta_{c,0}
     \Bigl\{ \delta_{b,0}\,\delta_{d,0} \theta^{\mu}
       + (\delta_{b,0}\,\delta_{d,1} - \delta_{b,1}\,\delta_{d,0})
         \theta^{\mu}\theta^{-\mu} \Bigr\}
    \nonumber\\
 && 
    {}+ \alpha\delta\, \delta_{a,0}\,\delta_{d,1}\Bigl\{
       \delta_{b,0}\,\delta_{c,0}
       - \delta_{b,0}\,\delta_{c,1}\theta^{\mu}
       + \delta_{b,1}\,\delta_{c,0}\theta^{-\mu}
       + \delta_{b,1}\,\delta_{c,1}\theta^{\mu}\theta^{-\mu}
    \Bigr\}
    \nonumber\\
 && 
    {}+ \beta\gamma\,\delta_{b,1}\,\delta_{c,0} \Bigl\{
       \delta_{a,0}\,\delta_{d,0}
       + \delta_{a,1}\,\delta_{d,0}\theta^{\mu}
       + \delta_{a,0}\,\delta_{d,1}\theta^{-\mu}
       + \delta_{a,1}\,\delta_{d,1}\theta^{\mu}\theta^{-\mu}
     \Bigr\}
    \nonumber\\
 && 
    {}- \beta\delta\,
     \delta_{a,0}\,\delta_{b,1}\,\delta_{c,0}\,\delta_{d,1}
     \theta^{\mu} \Tright{-2\mu}
    \nonumber\\
 && 
    {}+ \alpha\gamma \Bigl\{
     \delta_{b,0}\,\delta_{d,0}
      ( \delta_{a,1}\,\delta_{c,0} + \delta_{a,0}\,\delta_{c,1}\, )
    \nonumber\\*
 && \hspace{3em}
    {}+ \delta_{a,1}\,\delta_{c,1}\,\delta_{b,0}\,\delta_{d,0} \theta^{\mu}
    \nonumber\\*
 && \hspace{3em}
    {}+ \bigl\{
       \delta_{a,0}\,\delta_{c,1}
       ( \delta_{b,0}\,\delta_{d,1} + \delta_{b,1}\,\delta_{d,0} )
      + \delta_{a,1}\,\delta_{c,0}
       ( \delta_{b,0}\,\delta_{d,1} - \delta_{b,1}\,\delta_{d,0} )
        \bigr\} \theta^{-\mu}
     \nonumber\\*
 && \hspace{3em}
    {} + \delta_{a,1}\,\delta_{c,1}
       ( \delta_{b,1}\,\delta_{d,0} + \delta_{b,0}\,\delta_{d,1} )
       \theta^{\mu}\theta^{-\mu}
\Bigr\} \Tright{2\mu}
   \Bigr] \left( \theta^{K'} \vee \theta^{L'} \right), 
\end{eqnarray}
and 
\begin{eqnarray}
  \lefteqn{
 \left\{ (\theta^{\mu})^a (\theta^{-\mu})^b\theta^{K'}
  \vee
  \theta^{-\mu} \right\} \vee
    (\theta^{\mu})^c (\theta^{-\mu})^d\theta^{L'}
 }
 \nonumber\\
 &=& (\theta^{\mu})^a (\theta^{-\mu})^b\theta^{K'}
  \vee
  \left\{ \theta^{-\mu} \vee
    (\theta^{\mu})^c (\theta^{-\mu})^d\theta^{L'} \right\}
  \nonumber\\
 &=& (-)^{k'(c+d+1)} \Bigl[
      \delta_{b,0}\,\delta_{d,0}\Bigl\{
       \delta_{a,0}\,\delta_{c,0} \theta^{-\mu}
       + ( \delta_{a,1}\,\delta_{c,0} - \delta_{a,0}\,\delta_{c,1} )
         \theta^{\mu}\theta^{-\mu}
     \Bigr\} 
    \nonumber\\
 && 
     {}+ \alpha\delta\,\delta_{a,1}\,\delta_{d,0}\Bigl\{
      \delta_{b,0}\,\delta_{c,0}
      + \delta_{b,0}\,\delta_{c,1} \theta^{\mu}
      - \delta_{b,1}\,\delta_{c,0} \theta^{-\mu}
      + \delta_{b,1}\,\delta_{c,1} \theta^{\mu}\theta^{-\mu}
     \Bigr\}
   \nonumber\\
 && 
     {}+ \beta\gamma\, \delta_{b,0}\,\delta_{c,1} \Bigl\{
      \delta_{a,0}\,\delta_{d,0}
      + \delta_{a,1}\,\delta_{d,0}\theta^{\mu}
      + \delta_{a,0}\,\delta_{d,1}\theta^{-\mu}
      + \delta_{a,1}\,\delta_{d,1}\theta^{\mu}\theta^{-\mu}
     \Bigr\}
   \nonumber\\
 && 
    {}+ \beta\delta \Bigl\{
     \delta_{a,0}\,\delta_{c,0}
      ( \delta_{b,1}\,\delta_{d,0} + \delta_{b,0}\,\delta_{d,1} )
    \nonumber\\*
 && \hspace{3em}
     {}+ \bigl\{
      \delta_{b,0}\,\delta_{d,1}
       ( \delta_{a,1}\,\delta_{c,0} - \delta_{a,0}\,\delta_{c,1} )
      + \delta_{b,1}\,\delta_{d,0}
       ( \delta_{a,1}\,\delta_{c,0} + \delta_{a,0}\,\delta_{c,1} )
     \bigr\}\theta^{\mu}
   \nonumber\\*
 && \hspace{3em}
     {}+ \delta_{b,1}\,\delta_{d,1}\,\delta_{a,0}\,\delta_{c,0}
      \theta^{-\mu}
   \nonumber\\*
 && \hspace{3em}
     {}+ \delta_{b,1}\,\delta_{d,1}
        ( \delta_{a,1}\,\delta_{c,0} + \delta_{a,0}\,\delta_{c,1} )
       \theta^{\mu}\theta^{-\mu}
   \Bigr\} \Tright{-2\mu}
   \nonumber\\
 && 
    {}-\alpha\gamma \,
    \delta_{a,1}\,\delta_{b,0}\,\delta_{c,1}\,\delta_{d,0}
     \theta^{-\mu} \Tright{2\mu}
   \Bigr] \left( \theta^{K'} \vee \theta^{L'} \right).
\end{eqnarray}
These proofs compete the proof of the associativity of the particular form 
(\ref{eq:app-D-associativity-1form}) and thus lead to the associativity 
of the general form of Clifford product on the symmetric lattice.
\\

\noindent
{\Large \textbf{E. Adjoint operator of exterior derivative: 
$\delta=d-d\vee{}$}\\[0pt]}
\label{app:adjoint-op}

\renewcommand{\theequation}{E.\arabic{equation}}
\setcounter{equation}{0}

We show that the following operators,
\begin{eqnarray}
 \delta^{(1)}
   &=& d-d\vee{}
   = \frac{1}{2}\sum_\mu \left(
     \ddx{\theta^{-\mu}}\del_{-\mu} - \ddx{\theta^\mu}\del_{+\mu}\right)
   \equiv \sum_\mu \delta_\mu^{(1)},\\
 \delta^{(2)}
   &=& d-d\vee{}
   = -\sum_\mu \ddx{\theta^\mu}\del_{+\mu}
   \equiv \sum_\mu \delta_\mu^{(2)}, 
\end{eqnarray}
play the role of adjoint of exterior derivative 
\begin{equation}
d=\sum_\mu(\theta^\mu\del_{-\mu}-\theta^{-\mu}\del_{+\mu}) 
\equiv \sum_\mu d_\mu
\end{equation}
for the definitions of Clifford product given in (\ref{s-clifford-product}) 
and (\ref{another-symmetric-clifford-product}), respectively.
They satisfy 
\begin{eqnarray}
 \int \left( \zeta g\theta^K \right)^* \vee df \theta^L \Bigr|_\zform \vol
  = \int \left( \zeta  \delta^{(i)} g \theta^K \right)^*
     \vee f \theta^L \Bigr|_\zform \vol,
     \label{eq:delta-int}
\end{eqnarray}
for $i=1,2$.
More explicitly we give proofs of the following equivalence:
\begin{eqnarray}
 \left( \zeta g\theta^K \right)^* \vee  d_\mu f \theta^L \Bigr|_\zform
  = (\zeta \delta^{(i)}_\mu g \theta^K)^*
     \vee f \theta^L \Bigr|_\zform,
     \label{eq:delta-instead}
\end{eqnarray}
up to overall shift operators for $i=1,2$.
Summing up $\mu$ and integrating out, we obtain (\ref{eq:delta-int})
from (\ref{eq:delta-instead}) up to the surface terms.

Let us factor out $\theta^{\pm\mu}$ from $\theta^K$ and $\theta^L$,
\begin{eqnarray}
 \theta^K = (\theta^\mu)^a (\theta^{-\mu})^b \theta^{K'},
  & \qquad &
 \theta^L = (\theta^\mu)^c (\theta^{-\mu})^d \theta^{L'},
\end{eqnarray}
where $\theta^{K'}$ and $\theta^{L'}$ contain neither $\theta^\mu$ nor
$\theta^{-\mu}$. 
Then we can show the following factorization property:
\begin{eqnarray}
 \lefteqn{
 \left( \zeta 
      g' (\theta^\mu)^a (\theta^{-\mu})^b  \right)^* \vee 
     d_\mu f'
    (\theta^\mu)^c (\theta^{-\mu})^d 
      \Bigr|_\zform
     (\zeta \theta^{K'})^* \vee \theta^{L'} \Bigr|_\zform}
  \nonumber\\
  &=&
      (\zeta \delta_\mu^{(i)} g' (\theta^\mu)^a (\theta^{-\mu})^b)^*
     \vee f' (\theta^\mu)^c (\theta^{-\mu})^d \Bigr|_\zform
        (\zeta \theta^{K'})^* \vee \theta^{L'} \Bigr|_\zform
\end{eqnarray}
where $g' = T_{-K'} g$ and $f' = T_{-K'}f$ and those shift operators 
are generated by the relations, $(\theta^{K'})^* g = g'(\theta^{K'})^*$ and 
$(\theta^{K'})^* f = f'(\theta^{K'})^*$. 
Due to this factorization property we just need to prove the formula 
only for the factorized case. 

Introducing the most general form which include $\theta^{\pm\mu}$, 
\begin{eqnarray}
F&=& f_0+f_+\theta^\mu+f_-\theta^{-\mu} + \tilde{f} \theta^\mu \theta^{-\mu},\\
G&=& g_0+g_+\theta^\mu+g_-\theta^{-\mu} + \tilde{g} \theta^\mu \theta^{-\mu},
\end{eqnarray}
we need to prove
\begin{eqnarray}
(\zeta G)^* \vee d_\mu F \Bigr|_\zform = 
(\zeta\delta_\mu^{(i)} G)^* \vee F \Bigr|_\zform, 
\end{eqnarray}
up to overall shift operators.
Explicit calculations lead
\begin{eqnarray}
(\zeta G)^* \vee d_\mu F \Bigr|_\zform &=& 
(\zeta\delta_\mu^{(1)} G)^* \vee F \Bigr|_\zform  \nn \\
&=& \frac{1}{2} (\del_{+\mu}g_-^* - \del_{-\mu}g_+^*)f_0 + 
\frac{1}{4} \tilde{g}^*(\del_{-\mu}f_+ + \del_{+\mu} f_-),
\end{eqnarray}
and 
\begin{eqnarray}
(\zeta G)^* \vee d_\mu F \Bigr|_\zform = 
(\zeta\delta_\mu^{(2)} G)^* \vee F \Bigr|_\zform 
= -(\del_{-\mu}g_+^*) f_0.
\end{eqnarray}
Here we have used the lattice version of integration by part, 
$g(\del_{\pm\mu}f) = - (\del_{\mp\mu}g)f$ which is valid 
up to surface terms. 
We have thus completed the proof for the factorized version which leads 
the complete proof of the general form due to the factorization property.
\\


\end{document}